\documentclass[aps,superscriptaddress,showpacs,nofootinbib,eqsecnum,prd,notitlepage]{revtex4} 

\usepackage{epsf}  
\usepackage{eucal} 
\usepackage{amssymb}
\usepackage{amsmath} 
\usepackage{graphicx} 
\usepackage{bm}
\usepackage{dcolumn} 
\usepackage{epsfig} 
\usepackage{multirow}
\usepackage{hyperref}
\usepackage{color}

\begin{document}

\title{Power spectrum for inflation models with quantum and thermal noises}



\author{Rudnei O.  Ramos} \email{rudnei@uerj.br}
\affiliation{Departamento de F\'{\i}sica Te\'orica, Universidade do
  Estado do Rio de Janeiro, 20550-013 Rio de Janeiro, RJ, Brazil}

\author{L. A. da Silva} \email{las.leandro@gmail.com}
\affiliation{Departamento de F\'{\i}sica Te\'orica, Universidade do
  Estado do Rio de Janeiro, 20550-013 Rio de Janeiro, RJ, Brazil}


\begin{abstract}

We determine the power spectrum for inflation models covering all
regimes from cold (isentropic) to warm (nonisentropic) inflation.  We
work in the context of the stochastic inflation approach, which can
nicely describe both types of inflationary regimes concomitantly. A
throughout analysis is carried out to determine the allowed parameter
space for simple single field polynomial chaotic inflation models that
is consistent with the most recent cosmological data from the
nine-year Wilkinson Microwave Anisotropy Probe (WMAP) and in
conjunction with other observational cosmological sources.  We present
the results for both the amplitude of the power spectrum, the spectral
index and for the tensor to scalar curvature perturbation amplitude
ratio.  We briefly discuss cases when running is present.  Despite
single field polynomial-type inflaton potential models be strongly
disfavored, or even be already ruled out in their simplest versions in
the case of cold inflation, this is not the case for nonisentropic
inflation models in general (warm inflation in particular), though
higher order polynomial potentials (higher than quartic order) tend to
become  less favorable also in this case, presenting a much smaller
region  of parameter space compatible with the recent observational
cosmological data. 
Our findings also remain valid in face of the recently released Planck
results. 

\end{abstract}


\maketitle


\section{Introduction}
\label{sec1}

Microphysics during cosmological inflation is largely believed to
provide the  key ingredients responsible for seeding the large-scale
structure formation in the Universe. {}Fluctuations generated during
an early phase of inflation provide a source of nearly scale invariant
density perturbations. This prediction from inflation is found to be
nicely consistent with the cosmological observations, in particular,
with those from  the measurements of the Cosmic Microwave Background
(CMB) radiation~\cite{COBE,WMAP,WMAP9yr}. 

Primordial density perturbations originating during inflation can come
from both microscopic field dynamics and metric fluctuations. Those
from field fluctuations have been identified and previously been
studied as coming from two well separated and distinct sources,
according to the quantum field dynamics involved during inflation. The
details of inflation model building, like the field content and the
couplings of the inflaton field to other fields or to light field
degrees of freedom, can largely affect the dynamics during
inflation. {}For instance, they determine how ultimately the vacuum
energy density stored in the  inflaton field end up converted into
radiation, allowing the Universe to make the transition from the
inflationary to the radiation dominated phase. In one extreme case,
where the couplings of the inflaton field to other field degrees of
freedom can be neglected during inflation (but necessarily become
important later on in a preheating/reheating phase), density
fluctuations are mostly sourced by {\it quantum fluctuations} of the
inflaton field.  This is the case of the cold, or isentropic inflation
scenario~\cite{liddlebook}.  In cold inflation a thermal radiation
bath is only generated at the end of inflation, after the decay
products of the inflaton field produced during preheating/reheating
thermalize.  In the opposite case, it may happen that the couplings
among the various fields are sufficient strong to effectively generate
and keep a quasi-equilibrium thermal radiation bath throughout the
inflationary phase. In this situation, the inflationary phase can
smoothly connect to the radiation dominated epoch, without the need, a
priori, of a separate reheating period. This alternative inflationary
scenario was  named nonisentropic, or warm inflation (for recent
reviews, see, e.g., Refs.~\cite{WIreviews1,WIreviews2}). In warm
inflation  the primary source of density fluctuations comes from {\it
  thermal fluctuations}, which originate in the radiation bath and are
transferred to the inflaton~\cite{warmpert1,warmpert2,warmpert}.  

Thermal fluctuations in warm inflation can be modeled by a stochastic
Langevin equation (SLE), where a dissipative term describes the
transfer of  energy from the inflaton field to the thermal radiation
bath, while the backreaction of the thermal fluctuations in the
radiation bath  on the inflaton field is described by a stochastic
noise term.  This SLE can be derived from first
principles~\cite{GR,BGR,BR,BMR} and it has been shown~\cite{MX,GM,MBR}
to appropriately  govern the evolution of the inflaton perturbations
during warm inflation.  The formation and maintenance of a thermal
radiation bath during warm inflation is highly dependent on the
microscopic quantum field theory details involved, like the type and
magnitude of the interactions of the inflaton field with  the other
field degrees of freedom and how these other field degrees of freedom
interact among themselves. 

Specific model realizations of warm inflation and the details of the
quantum field theory involved have been reviewed recently in
\cite{WIreviews1,WIreviews2}. Despite the realization of the warm
inflation scenario be mostly a model dependent problem, the basic
conditions for warm inflation dominating over the cold inflation
scenario are fairly simple and can be expressed in terms of the
magnitude of the effective dissipation coefficient $\Upsilon$ in the
effective inflaton's equation of motion,   the temperature $T$ of the
radiation bath (assuming fast enough thermalization of the decay
products during warm inflation, which is again also a model dependent
problem) and on the Hubble scale during inflation, $H$.  Accounting
for a nonvanishing radiation bath, with energy density $\rho_R$, an
accelerated inflationary epoch in the Universe is only possible if
$\rho_R \ll \rho_\phi$, where $\rho_\phi$ is the inflaton energy
density.  Even if $\rho_R$ is much less than $\rho_\phi$, we can still
have  $\rho_R^{1/4} > H$. But since $\rho_R \sim T^4$, we typically
have for warm inflation that $T > H$.  This is the basic condition
that indicates that thermal  fluctuations should dominate over the
quantum inflaton field fluctuations, which are $\delta \varphi \sim
H$,  when computing the inflaton power spectrum in the density
perturbation calculation.  One would think that warm inflation would
also mean that the dissipation due to radiation production should also
dominate over the friction due to the metric expansion in the inflaton
equation of motion, $\Upsilon > 3 H$. But this only determines the
regime of strong dissipation in warm inflation, since in general we
can still have  dissipation as small as $\Upsilon \sim 10^{-7} H$,
like for example for an inflaton field with a quadratic
potential~\cite{HR}, or even for values $\Upsilon \gtrsim 10^{-12} H$,
when including a quartic potential for the inflaton~\cite{FangLee} and
still have a smooth transition from the inflationary phase to the
radiation  dominated epoch, one of the trademarks of warm inflation.

In nonisentropic inflation models, most notably in warm inflation,
there is thus a non negligible contribution from the radiation bath to
the power spectrum. With the dominant contribution to density
perturbations in warm inflation models coming  from  thermal
fluctuations. This is opposite to the more common cold inflation
models, where  the quantum fluctuations of the inflaton field make the
sole contribution. Little, or no  importance, has been given in the
literature so far to those cases where neither quantum or thermal
fluctuations make  the dominant  contribution to the  power spectrum,
with the possibility of fluctuations from both quantum and thermal
sources being equally important for the determination of the power
spectrum. In those situations, which in fact should cover most of the
parameter  space of realistic quantum  field theory models for
inflation, we are required to have a complete determination of both
types  of contributions to the power spectrum, coming from quantum and
thermal sources, and also to determine under which conditions one may
eventually dominate over the other. This is a  most required study,
since, as more precise measurements of the CMB radiation anisotropies
rapidly become available, greater demand is placed  on more precise
theoretical predictions, such that also better constraints on these
models can also be obtained, when contrasting them with the
observational data. This requires accounting for all important effects
contributing  to the density perturbations.  {}For warm inflation,
this requires knowing precisely when the eventual thermalized inflaton
fluctuations start effectively to dominate over the quantum ones and
also to be able to describe all the possible intermediate regimes
ranging from cold to warm inflation.  We here discuss  this
calculation of the total power spectrum for the inflaton, leading to a
result that should cover all the regimes ranging from cold to warm
inflation.

As mentioned above, in warm inflation the power spectrum from thermal
fluctuations is determined from a SLE for the inflaton field
fluctuations. {}For cold inflation  there is also an analogous
approach.  In the stochastic inflation program for cold inflation,
originally proposed by Starobinsky~\cite{Starobinskyoriginal}  (see
also~\cite{rey,LindeSI1,Sasaki,Starobinsky1994}),  the quantum
inflaton field is split in long-wavelength, super-Hubble modes and in
short-wavelength, sub-Hubble modes. While the super-Hubble component
effectively behaves like a classical system, it is influenced by the
sub-Hubble quantum modes, constituting the environment and that
behaves as a stochastic  noise term. This leads to a stochastic
equation, from which the effects of the quantum fluctuations can then
be taken into account. This represents an effective coarse-graining
performed on the quantum inflaton field and on a chosen scale that is
at least of the  de Sitter horizon length~$1/H$. 

In the warm inflation program, we start by integrating over field
degrees of freedom other than the inflaton field. The resulting
effective equation for the inflaton field turns out to be a
Langevin-like equation with dissipation and stochastic noise terms. We
can then further perform  now the coarse-graining on the inflaton
field according to the stochastic inflation program. This way one can
describe on equal footing the combined effects of both the  inflaton
quantum and thermal fluctuations (we will be assuming that a
thermalized radiation  bath exists, which, as mentioned above is a
model building problem).   By computing the inflaton field
perturbation power spectrum, we obtain a formula that reproduces the
cold and warm inflation results as limiting cases, but that can now
also describe the intermediate regimes between the two scenarios.  As
an extra bonus of our study, we are able to describe how quantum and
thermal components of the field fluctuations get entangled (a result
that may eventually be relevant to shed light on the decoherence
problem of cosmological perturbations during inflation, though we will
not address this specific theme in this work).

This paper is organized as follows. In section \ref{sec2} we briefly
review the construction leading to the expressions describing the
dynamics of the inflaton field in both the stochastic (cold) inflation
and in the nonisentropic (warm) inflation cases.  In section
\ref{sec3} the two programs for describing the quantum and thermal
fluctuations are combined and the relevant  equations are derived. In
section \ref{sec4} the inflaton field power spectrum for the
fluctuations is derived and studied.  It is briefly discussed the
cases when running is present.  In section \ref{sec5} we give our
results for simple polynomial inflaton field potentials for chaotic
inflation (which are the type of potentials that are presently under
stronger scrutiny and which are the most constrained by the
observational cosmological data) for both the amplitude for the power
spectrum, the spectral index and for the tensor to scalar curvature
perturbation amplitude ratio.  Our concluding remarks are given in
section \ref{sec6}. An appendix is included where we show some
technical details. 


\section{Langevin-like equations for cold and warm inflation}
\label{sec2}

Let us start by briefly reviewing the derivation of the inflaton's
equation  of motion in the context of the Starobinsky's stochastic
inflation approach and then, next, for the case of warm inflation.


\subsection{Stochastic approach for cold inflation}

Assuming that the couplings of the radiation fields to the inflaton
field are negligible during inflation, as usually assumed  in cold
inflation, then the equation of motion for the inflaton  field is the
standard one, 

\begin{equation} \label{coldeom}
\left[ \frac{\partial^2}{\partial t^2}  + 3H \frac{\partial}{\partial
    t} - \frac{1}{a^2}\nabla^2 \right] \Phi+  \frac{\partial
  V(\Phi)}{\partial \Phi} = 0 \;,
\end{equation}
where $\Phi$ here is the Heisenberg inflaton field operator,
$H=\dot{a}/a$ and $a$ is the scale factor. We will also restrict
ourselves to the case of the dynamics during the inflationary regime,
in which  case the metric is well approximated by the de Sitter one,
where $ds^2= dt^2-a^2(t) (d{\bf x})^2$ and $a(t)=\exp(Ht)$, with the
Hubble constant as given by the Friedmann equation.

The stochastic inflation approach is based on a coarse-graining of the
inflaton field, where the quantum inflaton field is decomposed in a
long wavelength part, $\Phi_>$, and in a short wavelength part,
$\Phi_<$,

\begin{equation} \label{decomposition}
\Phi({\bf x},t) \rightarrow \Phi_{>}({\bf x},t) + \Phi_{<}({\bf x},t)
\;, 
\end{equation}
and focus is given on the dynamics of the long-wavelength modes of the
field.  The short wavelength part of the field, $\Phi_<$, summarizes
quantum vacuum fluctuations, representing high momentum modes of the
field, with typically $k \gtrsim k_h \approx aH$.  This separation of
modes is implemented through a suitable filter function.  {}Following
the original works, e.g. Refs.
\cite{Starobinskyoriginal,rey,LindeSI1,Sasaki,Starobinsky1994},
$\Phi_<$ is written in terms of a filter (window) function, which
projects out the long wavelength modes and it is decomposed as

\begin{equation} \label{modeexpansion}
\Phi_<({\bf x},t) \equiv \phi_q({\bf x}, t) = \int \frac{d^3 k}{(2
  \pi)^{3/2}} W(k,t)\left[\phi_{\bf k}(t) e^{-i{\bf k} \cdot {\bf x}}
  \hat{a}_{\bf k} + \phi_{\bf k}^{*}(t) e^{i{\bf k} \cdot {\bf x}}
  \hat{a}^{\dagger}_{\bf k}\right] \;,
\end{equation}
where  $\phi_{\bf k}(t)$ are the field modes in momentum space,
$\hat{a}^{\dagger}_{\bf k}$ and $\hat{a}_{\bf k}$ are the creation and
annihilation operators,  respectively, and $W(k,t)$ is a suitable
window filter function.  The filter function $W(k,t)$ in the above
equation, in its most common implementation, like in the original
works on stochastic inflation, is expressed  in terms of a sharp
momentum cutoff,

\begin{equation} \label{window}
W(k,t) = \theta(k-\epsilon a H ) \;,
\end{equation}
where $\epsilon$ is a suitable small number, such that $\Phi_<({\bf
  x},t)$ refers only to the fields modes with wavelength much smaller
than the horizon.  ${\bf k}$ is the comoving coordinate wave-vector
and $k \equiv |{\bf k}|$. Overall, the results obtained from the
window filter Eq. (\ref{window}), like for the scalar perturbation
amplitude and other quantities, are found to be very weakly dependent
on the value of this parameter $\epsilon$, as we will also explicitly
see in our calculations to be presented in the next section.  It is
also known that a sharp window function, like  Eq.~(\ref{window}),
produces correlation functions  that tend to decay too slowly at large
distances, compared to  non-smoothed fields~\cite{Winitzki2000}.  The
use of a smooth window function adds a nonsharp smoothing scale and
can give the expected asymptotic behavior for the correlations.  {}For
example, a Gaussian function~\cite{Winitzki2000}, given by

\begin{equation} \label{gaussian}
W(k,t) = 1-\exp{\left[-\frac{k^2}{2(\epsilon aH)^2}  \right]}\;,
\end{equation}
has also been used in the literature. It works nicely as a smooth
window function, from which we can recover the sharp filter case as an
approximation, for modes with wavelengths sufficiently shorter than
the horizon  (or in the approximation $\epsilon \ll 1$).
 
In Eq. (\ref{modeexpansion}),  the field modes  $\phi_{\bf k}(t)$
satisfy

\begin{equation}\label{modeeq}
\left[\frac{\partial^2}{\partial t^2} + 3H \frac{\partial}{\partial t}
  + k^2/a^2 + \langle V_{,\phi\phi}(\Phi_>) \rangle \right] \phi_{\bf
  k}(t) = 0 \;,
\end{equation}
and

\begin{equation}
\label{moderel}
\phi_{\bf k}(t) \dot{\phi}^*_{\bf k}(t) -  \phi^*_{\bf k}(t)
\dot{\phi}_{\bf k}(t) = i/a^3 \;,
\end{equation}
where $\langle V_{,\phi\phi}(\Phi_>) \rangle$ is an average value of $
V_{,\phi\phi}(\Phi_>)$ that is taken with respect to the de Sitter
invariant vacuum. In its most general form, $\phi_{\bf k}(t)$ is given
by~\cite{books}

\begin{equation}
\phi_{\bf k}(t) = \frac{H\sqrt{\pi}}{2}\left(\frac{1}{aH}\right)^{3/2}
H_{\mu}^{(1)}\left(\frac{k}{aH}\right)\;,
\label{gensol}
\end{equation}
where $H_{\mu}^{(1)}(x)$ is the Hankel function of the first kind,
expressed in terms of the Bessel functions of the first and second
kind as

\begin{equation}
H_{\mu}^{(1)}(x) = J_\mu(x) + i Y_\mu(x)
\end{equation}
and $\mu = \sqrt{9/4 - m^2/H^2}$,  with $m^2 \equiv \langle
V_{,\phi\phi}(\Phi_>) \rangle \simeq {\rm constant}$.  {}For the high
frequency quantum part of the inflaton field, and when approximated as
a free massless field, $m^2=0$, the solution for the modes
$\phi_{{\bf k}}(t)$ that follows from Eq.~(\ref{gensol}), becomes

\begin{equation}\label{phi}
\phi_{{\bf k}}(t) \simeq \frac{H}{\sqrt{2k}} \left[ \tau -
  i\frac{1}{k}\right] e^{-ik\tau} \;,
\end{equation}
where $\tau =-1/[a(t)H]$ is the conformal de Sitter time (ranging from
$-\infty$ to 0).

{}Finally, using the field mode decomposition
Eq. (\ref{decomposition}) in Eq. (\ref{coldeom}), the equation of
motion for the long wavelength part of the quantum field, $\Phi_>$,
becomes

\begin{equation} \label{coldstoch}
\frac{\partial^2 \Phi_{>}}{\partial t^2}  + 3H \frac{\partial
  \Phi_{>}}{\partial t} - \frac{1}{a^2}\nabla^2 \Phi_{>} +
V_{,\phi}(\Phi_{>}) = \xi_q\;,
\end{equation}
where, by keeping only the linear terms in $\Phi_{<}$, the expression
for the term $\xi_q$ in Eq. (\ref{coldstoch}) is given by

\begin{equation} \label{xiq}
\xi_q = -\left[ \frac{\partial^2}{\partial t^2} +
  3H\frac{\partial}{\partial t} -  \frac{1}{a^2} \nabla^2 +
  V_{,\phi\phi}(\Phi_>) \right] \phi_q \;.
\end{equation}
Using the definition Eq. (\ref{modeexpansion}), it can be shown that
the comutator  for $\xi_q({\bf x},t)$ is

\begin{equation}
\left[\xi_q({\bf x},t)  , \xi_q({\bf x}',t') \right] = 0 \;,
\label{qcomut}
\end{equation} 
which suggests that $\xi_q$ behaves like a c-number.  It can then be
shown that $\xi_q$ can be interpreted as a (Gaussian white) stochastic
noise term, where $\langle \xi_{q} \rangle =0$ and with two-point
correlation function (in the vacuum), by using
Eqs.~(\ref{modeexpansion}) and (\ref{xiq}),  given by

\begin{equation} 
\langle \xi_{q}({\bf x},t) \xi_{q}({\bf x}',t')\rangle = \int
\frac{d^3 k}{(2\pi)^3} e^{i {\bf k}.({\bf x}-{\bf x}')} {\rm Re}\left[
  f_{\bf k}(t) f^*_{\bf k}(t')\right]\;,
\label{xiqxiq0}
\end{equation}
where $f_{\bf k}(t)$ is defined in terms of the filter function $W$
and the field modes in momentum space as

\begin{equation}
f_{\bf k}(t) = \left[\frac{\partial^2 W}{\partial t^2} +  3H
  \frac{\partial W}{\partial t} \right] \phi_{{\bf k}}(t) +  2
\frac{\partial W}{\partial t} \frac{\partial {\phi}_{{\bf
      k}}(t)}{\partial t}\;.
\label{fkt}
\end{equation}
Equation (\ref{coldstoch}) can then be interpreted as a stochastic
equation for the low-frequency, high wavelength field modes, whose
effect of the quantum (high-frequency) modes is felt through a
stochastic noise term $\xi_q$.


\subsection{Langevin equation for warm inflation}

The derivation of the effective equation of motion for the inflaton
field in the warm inflation case also involves a selection of a
relevant field mode, in which we are interested in the dynamics, while
the remaining degrees of freedom are taken as environmental ones,
similar to the cold stochastic inflation case. The main difference is
that now the relevant object is the inflaton field itself, either the
zero mode homogeneous mode, in case we are interested only in the
background dynamics, or the homogeneous mode plus the purely classical
thermal fluctuations, in case of the analysis involving the density
perturbation from warm inflation. The degrees of freedom that are
regarded as environment are any other fields in the original model
Lagrangian. In short (for a more throughout discussion and extended
formal details involving the quantum field theory derivation,  see,
e.g., the reviews \cite{WIreviews1,WIreviews2}, or the original papers
\cite{firstWI} about the warm inflation scenario), one starts with a
model Lagrangian density of the generic form,

\begin{equation}
{\cal L}[\Phi,X,Y] = {\cal L}[\Phi] + {\cal L}[X] + {\cal L}[Y] +
{\cal L}_{\rm int}[\Phi,X] + {\cal L}_{\rm int}[X,Y] \;,
\label{LPhiXY}
\end{equation}
where $\Phi$ is the inflaton field, $X$ are any field or degrees of
freedom  coupled directly to the inflaton field, while $Y$ can be any
other fields not necessarily coupled to the inflaton, but are coupled
to $X$. The ${\cal L}_{\rm int}[\Phi, X]$  and ${\cal L}_{\rm
  int}[X,Y]$ give the interaction among these fields.  The evolution
of the inflaton field can be properly determined in the context  of
the in-in, or the Schwinger closed-time path functional
formalism~\cite{calzetta+hu}.

In the closed-time path formalism the time integration is along a
contour in the complex time plane, going from $t=- \infty$ to  $+
\infty$ (forward branch) and then back to $t=-\infty$  (backwards
branch). Fields are then identified on each of the time branches like,
e.g., $\Phi_1$ and $\Phi_2$, respectively, and so on for the other
fields. Due to the duplication of field variables in this formalism,
one can now have four two-point Green function that can be
constructed with each of these fields, with the (1,1)-component
identified with the causal Feynman Green function, e.g.,
$G_{1,1}(x,x')= \langle \hat{T} \Phi_1(x) \Phi_1(x')\rangle$, where
$\hat{T}$ is the time ordering operator.  By functionally integrating
over the $X$ and $Y$ fields in Eq.~(\ref{LPhiXY}) in this formalism,
we obtain the corresponding effective action for the inflaton field
$\Phi$, from which its effective equation of motion can be
derived~\cite{WIreviews1}.  This procedure results typically in a
nonlocal effective equation of motion for the inflaton field $\Phi$.
This equation displays both dissipation and non-Markovian stochastic
noise terms  and it is, thus, a generalized Langevin-like equation of
motion~\cite{BMR}.  This is the expected behavior for the produced
dynamics of the system (the inflaton), when integrating over the
environmental degrees of freedom, here corresponding to the $X$ and
$Y$ fields (see for instance \cite{calzetta+hu} for a throughout
review and discussion about these type of equations and their
derivation in the context of quantum field theory).  


It is useful to consider a specific case as an example of this
procedure for showing how exactly a dissipation and noise terms can
emerge in a quantum field problem.  More specifically, let us consider
a model with a Lagrangian density of the form of Eq.~(\ref{LPhiXY}),  

\begin{eqnarray}
{\cal L} &=& \frac{1}{2} \left( \partial_\mu \Phi\right)^2 - V(\Phi)
\nonumber \\ &+& \frac{1}{2} \left( \partial_\mu X\right)^2
-\frac{1}{2} m_X^2 X^2 - \frac{g^2}{4} \Phi^2 X^2  \nonumber \\ &-&
{\cal L}_I[X,Y]\;,
\label{Lphichi}
\end{eqnarray}
where $X$ is a scalar field and  ${\cal L}[X,Y]$ denotes the part of
the Lagrangian density containing other fields $Y$ and their
interaction with $X$.  The $Y$ fields can for example represent
radiation fields, whose mass is much smaller than the temperature,
$m_Y \ll T$.  With $Y$ taken as another scalar field, we can choose a
simple form for ${\cal L}_I[X,Y]$, with a Yukawa coupling to $X$, for
example,

\begin{equation}
{\cal L}_I[X,Y] = h m_X X Y^2 + {\cal L}[Y]\;,
\end{equation}
whose specific form for the term involving only $Y$, ${\cal L}[Y]$,
will not be relevant for our derivation below.  The effective action
for the $\Phi$ field is determined by integrating out the fields $X$
and $Y$.  The functional integration over the $Y$ field will affect
the $X$ field two-point Green functions  $G^{(X)}_{ij}(x,x')$,
$i,j=1,2$,  dressing it by corrections coming from the $Y$ field.
Restricting to corrections to the $\Phi$ field up to quadratic order
in $X$, the functional  integration over the $X$ field in
Eq.~(\ref{Lphichi}) can then be exactly  performed. The procedure to
obtain the effective action for $\Phi$ and its corresponding effective
equation of motion follows analogous steps already shown in many
previous references (see, e.g.,  Refs.~\cite{WIreviews1,GR,BGR,BR,BMR}
for a more detailed account of this).  {}For the model Lagrangian
Eq.~(\ref{Lphichi}), keeping terms up to order ${\cal O}(g^4)$, which
is the order where nontrivial results first appear (e.g., dissipation
and fluctuation terms), the expression for the effective action for
$\Phi$, in the context of the  Schwinger closed-time path formalism,
becomes:

\begin{eqnarray}
\Gamma[\Phi_1,\Phi_2] &=&  \int d^4 x \left[\frac{1}{2} \left(
  \partial_\mu \Phi_1\right)^2 - V(\Phi_1) \right] -  \int d^4 x
\left[\frac{1}{2} \left( \partial_\mu  \Phi_2\right)^2 - V(\Phi_2)
  \right] + \nonumber \\ &-&\frac{g^2}{4}\int d^4x G^{(X)}_{11}(0)
\left[ \Phi_1^2(x)-\Phi_2^2(x)\right] \nonumber \\ &-& i
\frac{g^4}{16}\int d^4x d^4x' \left[ \Phi_1^2(x) G^{(X)}_{11}(x-x')^2
  \Phi_1^2(x') -  \Phi_2^2(x)G^{(X)}_{21}(x-x')^2\Phi_1^2(x')\right.
  \nonumber\\ &&\qquad
  \left. -\Phi_1^2(x)G^{(X)}_{12}(x-x')^2\Phi_2^2(x')
  +\Phi_2^2(x)G^{(X)}_{22}(x-x')^2\Phi_2^2(x')\right]\;.
\label{Gammaphi12}
\end{eqnarray}
At this point, it is convenient to define new field variables:
$\Phi_{c} = (\Phi_1 + \Phi_2)/2$ and $\Phi_{\Delta} =  \Phi_1 -
\Phi_2$, in terms of which the action (\ref{Gammaphi12}) can be
expressed in the form:

\begin{eqnarray}
\Gamma[\Phi_c,\Phi_\Delta] &=& \int d^4 x d^4 x' \left\{
\frac{i}{2}\Phi_\Delta (x) \Sigma_F[\Phi_c] (x,x')  \Phi_\Delta(x') +
\Phi_\Delta(x) {\cal O}_R[\Phi_{c}](x,x') \Phi_{c}(x')
\right. \nonumber \\ &+& \left. \frac{1}{4} \Phi_c(x) \Phi_\Delta(x)
\left[ -\frac{\lambda}{4!} \, \delta^4(x-x')+   \Sigma^{(X)}_R (x,x')
  \right] \Phi_\Delta^2(x') \right\} \;,
\label{GammaphicD}
\end{eqnarray}
where $\Sigma^{(X)}_{F}[\Phi_c](x,x') = \Phi_c(x) \Phi_c(x')
\Sigma^{(X)}_{F}(x,x')$, with

\begin{equation}
\Sigma_F(x,x')  = - \frac{g^4}{2}  \left[ F^{(X)} (x-x')^2 -
  \frac{1}{4} \rho^{(X)} (x-x')^2 \right] \;,
\label{SKchi}
\end{equation}
and

\begin{equation}
{\cal O}_R[\Phi_{c}](x,x') = \left[\partial^2 + V''(\Phi_c) +
  \Sigma^{(X)}_{R,{\rm local}}\right]  \delta^4(x-x') +
\Sigma^{(X)}_{R}[\Phi_c](x,x')\;,
\label{ORchi}
\end{equation}
where $\Sigma^{(X)}_{R}[\Phi_c](x,x') = \Phi_c(x) \Phi_c(x')
\Sigma^{(X)}_{R}(x,x')$, with

\begin{equation}
\Sigma^{(X)}_{R} (x,x') = - \frac{g^4}{2}  F^{(X)} (x-x') \rho^{(X)}
(x-x') \, \theta(t-t')\;,
\label{SRchi}
\end{equation}
where the functions $F^{(X)}(x,x')$ and $\rho^{(X)}(x,x')$ in
Eqs.~(\ref{SKchi}) and (\ref{SRchi}), are defined  in terms of the 
expectation value for the commutator and the anticommutator of the
$X$ field, respectively, by 

\begin{eqnarray}
\rho^{(X)}(x,x')&=& i\langle [X(x),X(x')]\rangle\;,
\label{rho}\\
F^{(X)}(x,x')&=&\frac{1}{2}\langle \{X(x),X(x')\}\rangle\;,
\label{functionF}
\end{eqnarray}
and in Eq.~(\ref{ORchi}),   $\Sigma^{(X)}_{R,{\rm local}} = -(g^2/2)
G_{11}^{(X)}(0)$ is a local correction to the $\Phi$ field action
coming from the integration over the $X$ field.

The first term in Eq. (\ref{GammaphicD}), the quadratic term in
$\Phi_\Delta$, is a purely imaginary term. It can be interpreted as
follows. We can perform  a Hubbard--Stratonovich transformation in the
functional partition function generating the effective action
(\ref{GammaphicD}) (see, e.g.. Ref.~\cite{GR}), introducing a
stochastic field $\xi(x)$ to decouple the quadratic term in
$\Phi_\Delta$,

\begin{eqnarray}
&& \exp \left\{ -\frac{1}{2} \int d^4 x d^4 x'\,  \Phi_\Delta (x)
  \Sigma_F^{(X)}[\Phi_c] (x,x') \Phi_\Delta(x') \right\} \nonumber
  \\ && = \int D \xi \; \exp \left\{ -\frac{1}{2} \int d^4 x d^4 x'\,
  \xi (x) \left[ \Sigma_F^{(X)}\right]^{-1}(x,x') \xi (x') + i \int
  d^4 x \xi(x) \Phi_\Delta(x) \right\}\;.
\label{noise}
\end{eqnarray}
The effective action expressed in terms of $\xi(x)$ has now only real
terms.  Next, the effective equation of motion for $\Phi$ is obtained
from the saddle point equation~\cite{GR}

\begin{equation}
\frac{\delta \Gamma[\Phi_c,\Phi_\Delta,\xi] }{\delta \phi_{\Delta} }
\Bigr|_{\phi_\Delta=0} = 0\;,
\label{eom}
\end{equation}
which, by using the effective  action written in terms of $\xi(x)$,
gives a stochastic equation of motion,

\begin{eqnarray}
\int d^4 x' {\cal O}_R[\Phi_{c}](x,x') \Phi_{c}(x') = \xi(x) \;,
\label{langevin}
\end{eqnarray}
where ${\cal O}_R[\Phi_{c}](x,x')$ is defined in Eq.~(\ref{ORchi}).
The equation (\ref{langevin}) can be seen as a nonlocal Langevin-like
equation of motion, where $\xi(x)$ can be interpreted as  a Gaussian
stochastic noise with the general properties of having zero mean,
$\langle \xi(x) \rangle =0$, and two-point correlation

\begin{equation}
\langle \xi(x) \xi(x') \rangle = \Sigma_F^{(X)}[\Phi_c] (x,x')\;.
\label{noisetwopoint}
\end{equation}

The equation of motion (\ref{langevin}) can be further put in a more
suitable form, similar to a non Markovian Langevin equation, if we
define $\Sigma_{R}^{(X)} (x,x') = \Sigma^{(X)}_\rho(x,x')
\theta(t-t')$ and a dissipation kernel ${\cal D}^{(X)}(x,x')$ as
\cite{BMR}

\begin{equation}
\Sigma^{(X)}_\rho(x,x') {\rm sgn}(t-t') = - \frac{\partial}{\partial
  t'} {\cal D}^{(X)} (x,x')\;.
\label{diss kernel}
\end{equation}
Then, by absorbing local terms in a renormalized effective potential
for $\Phi$, $V_{\rm eff,r}(\Phi)$, Eq. (\ref{langevin}) can now be
expressed as 

\begin{eqnarray}
\partial^2 \Phi_c + V_{\rm eff,r}'(\Phi_c) + \int d^4 x' {\cal
  D}^{(X)} (x,x')  \dot{\Phi}_c(x')= \xi(x) \; .
\label{langevin2}
\end{eqnarray}
The dissipation kernel ${\cal D}^{(X)} (x,x')$ in the above equation
and the noise kernel given by Eq.~(\ref{noise}) are related to each
order, under a space-time Fourier transform, by the
fluctuation-dissipation relation

\begin{equation}
\Sigma_F^{(X)}({\bf p},\omega) =  2\omega \left[ n(\omega) +
  \frac{1}{2} \right] {\cal D}^{(X)}({\bf p}, \omega)\;,
\label{chifdr}
\end{equation}
where $n(\omega)$ is the Bose-Einstein distribution.

The resulting nonlocal, non-Markovian equation of motion for $\Phi$,
Eq.~(\ref{langevin2}), can be shown to be well approximated by a local
Langevin equation of motion when there is a clear separation of
timescales in the system, which leads, in this local approximation, to
a local dissipation coefficient defined  by

\begin{equation}
\Upsilon= \int d^4 x' \, \Sigma_R[\Phi_c](x,x') \, (t'-t) \;,
\label{Upsilon}
\end{equation}
where,  for this specific example we have worked out here, with
Lagrangian density (\ref{Lphichi}), we have that

\begin{equation}
\Sigma_R[\Phi_c](x,x') = -i g^4 \Phi_c^2 \; \theta(t-t')
\langle[X^2(x),X^2(x')]\rangle\;.
\end{equation}
Complete expressions for $\Upsilon$ can be found in
Ref.~\cite{dissipcoefs} for different interactions and regimes of
parameters.  

The validity of the localization approximation for the nonlocal terms
in Eq.~(\ref{langevin2}) and region of parameters where this  can be
achieved in specific models, were discussed at length in many previous
references about the warm inflation implementation and we will not
enter or need these formal details here (for the interested reader,
please see, e.g., Refs.~\cite{BMR,WIreviews1} and references there
in).  In short, if the self-energies terms coming from the $X$ and $Y$
fields introduce a response  timescale $1/\Gamma$ and if $\Phi$ is
slowly varying on the response timescale $1/\Gamma$, 

\begin{equation}
\label{adiabatica}
\frac{\dot \Phi}{\Phi} \ll \Gamma,
\end{equation}
which is typically referred to as the adiabatic approximation, then a
simple Taylor expansion of the nonlocal terms can be performed.  The
typical microscopic mechanism for generating radiation in warm
inflation is through the transfer of energy from  $\Phi$ through $X$
that then can decay into the radiation fields $Y$.  Then, if $\Gamma >
H$, this radiation is thermalized sufficiently fast and, by further
requiring that $\Gamma  \gg \dot{T}/T$, it can be maintained in a
close to thermal equilibrium state.  Under these conditions, we can
write for the inflaton an effective SLE of motion (dropping the
subscript "c" in the field in Eq.~(\ref{langevin2}) and working
directly in a Friedmann-Robertson-Walker (FRW)  background
metric~\cite{BR})  of the form

\begin{equation}
\left[ \frac{\partial^2}{\partial t^2}  + (3H + \Upsilon)
  \frac{\partial}{\partial t} - \frac{1}{a^2}\nabla^2 \right] \Phi+
\frac{\partial V_{\rm eff,r}(\Phi)}{\partial \Phi} = \xi_T\;,
\label{eomWI}
\end{equation}
where $V_{\rm eff,r}(\Phi)$ is a renormalized effective potential for
the inflaton (after integrating over the $X$ and $Y$ fields),
$\Upsilon$ is the dissipation coefficient and $\xi_T$ describes
thermal (Gaussian and white) noise fluctuations in the local
approximation for the effective equation of motion and it  is
connected to the dissipation coefficient through a Markovian
fluctuation-dissipation relation, which in expanding space-time, is
given by

\begin{equation}
\langle \xi_T ({\bf x},t) \xi_T({\bf x}',t') \rangle =  2 \Upsilon T
a^{-3} \delta({\bf x}- {\bf x}') \delta(t-t')\;,
\label{flucdiss}
\end{equation} 
where the average here is to be interpreted as been taken over the
statistical ensemble.

The dissipation coefficient $\Upsilon$ in Eq. (\ref{eomWI}) depends on
the specifics of the model Lagrangian parameters entering in
Eq. (\ref{LPhiXY}) and can be a function of temperature and inflaton
field. Recently, it was derived in Ref. \cite{dissipcoefs} the
different dissipation coefficients arising from many different types
of couplings and for different temperature regimes that can be found
in warm inflation, with other relevant regimes for warm inflation also
identified in Ref.~\cite{dissipation}.  {}For example, for the models
treated in the  Refs.~\cite{WIreviews1,dissipcoefs}, the dissipation
coefficient  $\Upsilon$ was found to  have the following generic
dependence on the inflaton field amplitude $\phi$, temperature $T$ and
on the mass $m_X$ for the $X$ fields coupled to  the inflaton, 

\begin{equation}
\Upsilon = C_\phi \frac{T^c \phi^{2a}}{m_X^{2b}},\;\;\; c+2a-2b=1\,,
\end{equation}
where the values of the constants $a,b,c$  depend on the specifics of
the model construction for warm inflation and on the temperature
regime of the thermal bath.  {}For example, for a trilinear
interaction of the  inflaton field with a bosonic $X$ field and at low
temperature regimes  (for the temperature taken with  respect to the
intermediate field $X$ in the Lagrangian density model
Eq.~(\ref{LPhiXY}), $T < m_X$), it is found that $a=c=3,\,b=4$, while
for a quadratic coupling of the inflaton field  with a bosonic $X$
field like the one taken in the example shown above   and at high
temperatures ($T > m_X$), it is found that $a=1,\,b=0,\,c=-1$. Other
types of dependence of the dissipation coefficient on the temperature
and on the inflaton amplitude can be found for other forms of
couplings and regimes of parameters.

{}From a model building perspective, warm inflation has several
advantages.  {}For example, we can implement Lagrangian densities of
the form  (\ref{LPhiXY}) in terms of supersymmetry (SUSY)
\cite{WIreviews1,WIreviews2,BR1},  so couplings of the inflaton field
with the other (bosonic and fermionic)  fields $X$ and between $X$ and
$Y$ can be made fairly large, ${\cal O}(0.1)$,  and we can still keep
quantum corrections to the inflaton potential small  using
SUSY. {}Furthermore, temperature corrections to the inflaton potential
can as well be made small when $T < m_X$.


\section{Stochastic approach for cold and warm inflation}
\label{sec3}

We now want to combine both the Starobinsky stochastic inflation
approach  and the warm inflation scenario in a single SLE
description. Note that this could be implemented from the very
beginning in the warm inflation construction of Eq. (\ref{eomWI}),
where besides of integrating over the $X$ and $Y$ fields, a functional
integration of the sub-horizon modes of the inflaton field $\Phi$
could also be performed, leaving only an effective equation describing
the super-horizon part of $\Phi$. A functional integration of the
sub-horizon modes of the inflaton (in the absence of any other fields
coupled to the inflaton) and the construction of the influence
functional for the super-horizon modes, in the context of the cold
inflation only, has been done, for example,
in~\cite{inflfunct1,inflfunct2,inflfunct3}.  This functional
integration procedure, when applied to cold inflation, is able to
describe the inflaton dynamics in a more rigorous way  than the simple
field split implemented in the Starobinsky program and it has been
shown to  lead to additional contributions in the inflaton equation of
motion for  the long wavelength modes,  Eq. (\ref{coldstoch}).
However, it has also been shown    that these extra contributions only
produce small effects and they can, therefore, be
neglected~\cite{inflfunct1,inflfunct3}. Therefore, the result given by
Eq. (\ref{coldstoch}) can still be used. 

By describing the influence of the thermal radiation bath on the
inflaton  dynamics, according to the SLE (\ref{eomWI}), we can now
account concomitantly also for the effect of the  sub-horizon quantum
fluctuations on the dynamics for the long wavelength modes of the
inflaton field.  Hence, the inflaton field $\Phi$ in Eq.~(\ref{eomWI})
is again split like in Eq.~(\ref{decomposition}),  in a sub-horizon,
quantum fluctuation part, $\phi_q({\bf x},t)$, which can again be
described by Eq. (\ref{modeexpansion}), and in a super-horizon part,
that behaves essentially as the classical inflaton mode. The classical
inflaton mode is in turn split in the homogeneous inflaton field
$\phi(t)$ and in the fluctuations $\delta \varphi({\bf x},t)$.  The
background homogeneous inflaton $\phi(t)$ can be seen as the
coarse-grained inflaton field over approximately all de Sitter horizon
size $\Omega = 1/H$,

\begin{equation} \label{coarsegraining}
\phi(t) = \frac{1}{\Omega} \int_{\Omega}d^3{\bf x}  \Phi({\bf x},t) \;
,
\end{equation}
and $\delta \varphi ({\bf x},t)$ are  the classical fluctuations on
top of it. 

In the absence of the thermal backreaction from the radiation bath, or
when $T \ll H$, it is the quantum noise  $\xi_q$ that is expected to
predominantly contribute to $\delta \varphi({\bf x},t)$. Here, we will
not attempt to study how exactly the quantum fluctuations turn into
classical ones, the decoherence problem of the fluctuations in cold
inflation, but assume that the decoherence time is short enough that
when the inflaton fluctuations cross the horizon they behave or are
essentially classical fluctuations  already, as it is generically
expected (the decoherence problem  in inflation was also extensively
studied in the context of stochastic inflation and the influence
functional method. {}For some references, see, e.g.,
\cite{calzetta+hu,inflfunct1,decoherence} and references there in).
In addition, as the temperature of the radiation bath increases, or
when the coupling of the inflaton field with the radiation bath
becomes sufficiently strong, we expect that now the backreaction  of
the thermal radiation, represented by the thermal stochastic
fluctuations in Eq. (\ref{eomWI}), will become more and more
important, till it eventually surmounts that from the quantum
stochastic fluctuations. Next, we will study  how exactly these two
effects compete in giving the dominant contribution to the inflaton
fluctuations $\delta \varphi({\bf x},t)$. Towards this goal, and as
explained above, we can then write the inflaton field  as

\begin{equation}
\Phi({\bf x},t) = \phi(t) + \delta \varphi({\bf x},t) + \phi_q({\bf
  x},t)\;.
\label{splitTq}
\end{equation}
By substituting Eq. (\ref{splitTq}) in Eq. (\ref{eomWI}), and
expanding in both perturbations $\delta \varphi({\bf x},t)$ and
$\phi_q({\bf x},t)$ to first order, we obtain the equations for the
background field $\phi(t)$ and for the fluctuations $\delta
\varphi({\bf x},t)$, given, respectively, by

\begin{eqnarray}  
&&\frac{\partial^2 \phi}{\partial t^2}  + \left[3H + \Upsilon(\phi)
    \right]\frac{\partial \phi}{\partial t} + V_{,\phi}(\phi)  = 0\;, 
\label{back} \\  
&&\left\{\frac{\partial^2}{\partial t^2}  + \left[3H + \Upsilon(\phi)
  \right]\frac{\partial}{\partial t}  - \frac{1}{a^2}\nabla^2  +
\Upsilon_{,\phi}(\phi) \dot{\phi}+ V_{,\phi\phi}(\phi) \right\} \delta
\varphi = \tilde{\xi}_q + \xi_T \;,  
\label{deltavarphi} 
\end{eqnarray}
where $\tilde{\xi}_q$ is the quantum noise term Eq. (\ref{xiq}),  but
including now also the  contributions coming from the dissipation
coefficient due to the integration over any of the radiation bath
fields (for example, the fields $X$ and $Y$ in  Eq. (\ref{eomWI})),

\begin{equation}
\tilde{\xi}_q =  -\left\{ \frac{\partial^2}{\partial t^2} + \left[3H +
  \Upsilon(\phi)\right] \frac{\partial}{\partial t}  - \frac{1}{a^2}
\nabla^2+  \Upsilon_{,\phi}(\phi) \dot{\phi} + V_{,\phi\phi}(\phi)
\right\} \phi_q \;.
\label{qtilde}
\end{equation}
We note that we will be working always in the local approximation for
the SLE, in which case, both noise terms in Eq.~(\ref{deltavarphi})
are Markovian stochastic noises. We should also note  that in this
case, a quantum noise term that we would expect to be produced
alongside the thermal noise $\xi_T$, from the functional integration
of the the fields $X$ and $Y$, vanishes identically in this local
approximation~\cite{BMR} and the quantum noise term $\tilde{\xi}_q$
comes exclusively from the  inflaton field quantum modes, analogously
as in the standard stochastic inflation program (for a study of the
effects on the inflation dynamics of a quantum noise  in the
non-Markovian case see, e.g., Refs.~\cite{ng1,ng2}). 

In terms of the modified quantum noise $\tilde{\xi}_q$, it can be
verified that the commutator (\ref{qcomut}) still satisfies

\begin{equation}
\left[\tilde{\xi}_q({\bf x},t)  , \tilde{\xi}_q({\bf x}',t') \right] =
0 \;,
\label{qtildecomut}
\end{equation}
which again is suggestive of interpreting $\tilde{\xi}_q$ as a
classical quantity.

{}From the equation of motion for the background field $\phi(t)$,
Eq. (\ref{back}), we have the slow-roll equation as that of warm
inflation,

\begin{equation}
3H(1+Q)\dot \phi + V_{,\phi}(\phi) = 0\;,
\label{varphislow}
\end{equation}
where we have defined $Q = \Upsilon/3H$, and we can also define the
slow-roll  coefficients~\cite{warmpert2}

\begin{equation}
\varepsilon = \frac{1}{16\pi G}\left[ \frac{V_{,\phi}}{V}\right]^2 \ll
1+Q \;,
 \label{sra1}
\end{equation}

\begin{equation}
\eta = \frac{1}{8\pi G}\frac{V_{,\phi\phi}}{V} \ll 1+Q\;,
\label{sra2}
\end{equation}
and
\begin{equation}
\beta = \frac{1}{8\pi G}\frac{\Upsilon_{,\phi} V_{,\phi}}{\Upsilon V}
\ll 1+Q \;,
\label{sra3}
\end{equation}
where the third slow-roll coefficient, $\beta$, is typical of warm
inflation, accounting for the possible dependence of the dissipation
coefficient on the inflaton's amplitude.

Going to momentum space, it becomes  convenient to express the
equation for the fluctuations, Eq. (\ref{deltavarphi}), in term of the
variable  $z \equiv k/(aH)$.  If we also use the definitions of the
slow-roll  coefficients,  Eqs. (\ref{sra1}), (\ref{sra2}) and
(\ref{sra3}), the equation satisfied by the fluctuations $\delta
\varphi({\bf k},z)$, the Eq. (\ref{deltavarphi}) in momentum space and
in terms of the variable $z$,  can be written in the form

\begin{eqnarray} 
&&\delta \varphi''({\bf k},z) - \frac{1}{z}(3Q+2) \delta \varphi'({\bf
    k},z) + \left[1 + 3 \frac{\eta - \beta Q/(1+Q)}{z^2}\right] \delta
  \varphi({\bf k},z)   \nonumber \\ && = \frac{1}{H^2z^2}\left[
    \xi_T({\bf k},z) + \tilde{\xi}_q({\bf k},z) \right] \;,
\label{fluctk2}
\end{eqnarray}
where primes in $\varphi({\bf k},z)$ mean now derivatives with respect
to the variable $z$.  Equation (\ref{fluctk2}) for the inflaton
fluctuations generalizes the one obtained in \cite{GM}. Its general
solution can be expressed in terms of a Green function,

\begin{equation} 
\delta \varphi({\bf k},z) = \int_{z}^{\infty} dz'
G(z,z')\frac{(z')^{1-2\nu}}{z'^2 H^2} \left[\tilde{\xi}_q(z') +
  \xi_T(z')\right] \;,
\label{solutionbessel}
\end{equation}
where the Green function $G(z,z')$ is given in terms of Bessel
functions,

\begin{equation}
G(z,z') = \frac{\pi}{2}z^{\nu}z'^{\nu}\left[
  J_{\alpha}(z)Y_{\alpha}(z') - J_{\alpha}(z')Y_{\alpha}(z)\right] \;, 
\label{Gzz}
\end{equation}
with $z'>z$ and

\begin{eqnarray}
&&\nu = 3(1 + Q)/2\;,\nonumber\\  && \alpha = \sqrt{ \nu^2 + \frac{3
      \beta Q}{1+Q} - 3 \eta} \;.
\label{nualpha}
\end{eqnarray}

Using Eq. (\ref{solutionbessel}), we can write the following
expression for the correlation $\langle \delta \varphi({\bf k}',z)
\delta \varphi({\bf k}',z) \rangle $,

\begin{eqnarray} \label{powercorrelation}
\langle \delta \varphi ({\bf k},z) \delta \varphi ({\bf k}',z)\rangle
&=& \frac{1}{H^4}\int_{z}^{\infty} dz_2 \int_{z}^{\infty} dz_1
G(z,z_1)G(z,z_2)\frac{(z_1)^{1-2\nu}}{z_1^2}\frac{(z_2)^{1-2\nu}}{z_2^2}
\langle \tilde{\xi}_q({\bf k},z_1) \tilde{\xi}_q({\bf k}',z_2)\rangle
\nonumber \\  &+&   \frac{1}{H^4}\int_{z}^{\infty} dz_2
\int_{z}^{\infty} dz_1
G(z,z_1)G(z,z_2)\frac{(z_1)^{1-2\nu}}{z_1^2}\frac{(z_2)^{1-2\nu}}{z_2^2}
\langle \xi_T({\bf k},z_1) \xi_T({\bf k}',z_2)\rangle \;.
\end{eqnarray}
Note that there is no inter-correlation between the two stochastic
noise terms,  the thermal one $\xi_T$ and the quantum one
$\tilde{\xi}_q$, since  they come from different sources, as we have
already  discussed before.   Below, we evaluate the two terms in the
right-hand-side of Eq. (\ref{powercorrelation}) separately.


\section{Total power spectrum including quantum and thermal noises}
\label{sec4}

Let us now compute in details each one of the contributions to the
power spectrum appearing in Eq.~(\ref{powercorrelation}).


\subsection{The quantum noise component for the power spectrum}

Let us now evaluate the quantum noise component for the power
spectrum, which comes from the first term in right-hand-side of
Eq. (\ref{powercorrelation}). Here, we will follow a similar approach
as used in \cite{riotto2} (see also \cite{calzetta+hu}) for the
computation of the power spectrum in the stochastic cold inflation
case.  We first need to compute the quantum noise two-point
correlation, $\langle \tilde{\xi}_q({\bf k},z_1) \tilde{\xi}_q({\bf
  k}',z_2)\rangle$.  This is determined by the equation defining the
quantum noise $\tilde{\xi}_q$,  Eq. (\ref{qtilde}). 

{}From the {}Fourier transform  of Eq. (\ref{qtilde}) and working in
terms of the comoving time $\tau \equiv -1/(a H)$,  which
significantly facilitates the calculations, we find 

\begin{equation} \label{noiseFourierTau}
\tilde{\xi}_q({\bf k},\tau) = -H^2 \tau^2 \left[
  \frac{\partial^2}{\partial \tau^2}  -\frac{2+3Q}{ \tau}
  \frac{\partial}{\partial \tau} + k^2 + \frac{\Upsilon_{,\phi}(\phi)
    \dot \phi}{H^2 \tau^2}  + \frac{V_{,\phi\phi}(\phi)}{H^2 \tau^2}
  \right] \hat{\phi}_q({\bf k},\tau) \;,
\end{equation}
where

\begin{equation} \label{modeexpansionFourierTau}
\hat{\phi}_q({\bf k},\tau) =   W(k,\tau) \left[ \phi_{{\bf k}}(\tau)
  \hat{a}_{-{\bf k}} + \phi^*_{{\bf k}}(\tau) \hat{a}^{\dagger}_{\bf
    k} \right] \;.
\end{equation}
Under the assumption that the quantum fluctuations are \cite{Sasaki}
${\cal O}(\sqrt{\hbar})$ and also considering that the dissipation
term is at least of order \cite{BMR} ${\cal O}(\hbar)$, then the
quantum field modes  $\phi_{\bf k}$ still satisfy the usual
differential equation (in conformal time) at leading order:
 
\begin{equation} 
\left[ \frac{\partial^2}{\partial \tau^2}  -\frac{2}{ \tau}
  \frac{\partial}{\partial \tau} + k^2  +
  \frac{V_{,\phi\phi}(\phi)}{H^2 \tau^2}   \right] \phi_{\bf k}(\tau)
=0\;,
\label{modeeqUpsilon}
\end{equation}
with solution of the form of Eq. (\ref{gensol}), which in conformal
time is

\begin{equation}
\phi_{\bf k}(\tau) = \frac{H\sqrt{\pi}}{2} (|\tau|)^{3/2}
H_{\mu}^{(1)}\left(k |\tau|\right)\;,
\label{gensolQ}
\end{equation}
where $\mu = \sqrt{9/4 - 3 \eta} \approx 3/2 - \eta$.  

{}From Eqs. (\ref{noiseFourierTau}) and
(\ref{modeexpansionFourierTau}),  we obtain for the two-point
correlation for the quantum noise the expression

\begin{equation} 
\langle \tilde{\xi}_{q}({\bf k},\tau) \tilde{\xi}_{q}({\bf
  k}',\tau')\rangle = (\tau \tau')^2  H^4  \left[ f_{\bf k}(\tau)
  f^*_{\bf k'}(\tau') \langle \hat{a}_{-\bf k} \hat{a}^\dagger_{\bf
    k'} \rangle +  f^*_{\bf k}(\tau) f_{\bf k'}(\tau') \langle
  \hat{a}^\dagger_{\bf k'} \hat{a}_{-\bf k} \rangle \right]\;,
\label{xiqxiq}
\end{equation}
where $f_{\bf k}(\tau)$ is defined as

\begin{equation}
f_{\bf k}(\tau) = \left[W'' - \frac{2+3Q}{\tau} W' - \frac{3 \beta
    Q}{(1+Q)\tau^2}W  \right] \phi_{{\bf k}}(\tau) +  2 W'
{\phi'}_{{\bf k}}(\tau)\;.
\label{fktau}
\end{equation}

In Eq.~(\ref{xiqxiq}), the averages involving the creation and
destruction operators,  $\langle \hat{a}_{-{\bf k}}
\hat{a}^{\dagger}_{{\bf k}'} \rangle$ and  $\langle
\hat{a}^\dagger_{\bf k'} \hat{a}_{-\bf k} \rangle$, are evaluated in a
thermal radiation bath, so they are in fact quantum statistical
averages.  We use the definitions, $\langle \hat{a}_{-{\bf k}}
\hat{a}^{\dagger}_{{\bf k}'} \rangle = [n(k) +1] (2 \pi)^3 \delta({\bf
  k} + {\bf k}')$ and $\langle \hat{a}^\dagger_{\bf k'} \hat{a}_{-\bf
  k} \rangle =  n(k) (2 \pi)^3 \delta({\bf k} + {\bf k}')$, where
$n(k)$ is the distribution function for the  high-frequency quantum
inflaton modes. In warm inflation, it assumes the thermal Bose
distribution form, $n(k) = 1/[\exp(k/(a T))-1]$. It is been implicitly
assumed here  that the  inflaton fluctuation momentum modes crossing
the horizon have had enough time to  reach thermal equilibrium with
the radiation bath. This condition of thermal equilibrium for the
inflaton fluctuations is mostly a model dependent problem, requiring
the study of the dynamical  microphysics involved,  the details of the
interactions involved and the knowledge of the field content and
properties of the radiation bath. However, it can be verified and
checked that thermal equilibrium can be achieved in general through
the study of the Boltzmann dynamics for the particle's distributions,
which can be obtained from the determination of the rate of energy
transfer from the background inflaton field to the radiation bath and
whose formal details can be found, e.g., in Ref.~\cite{dissGraham} and
also discussed in Ref.~\cite{dissipation}.

The expression (\ref{xiqxiq}) for the two-point correlation for the
quantum noise can then be written as

\begin{equation} 
\langle \tilde{\xi}_{q}({\bf k},\tau) \tilde{\xi}_{q}({\bf
  k}',\tau')\rangle = (2 \pi)^3 \delta({\bf k} + {\bf k}') (\tau
\tau')^2  H^4  \left[2n(k) +1 \right]  {\rm Re}\left[ f_{\bf k}(\tau)
  f^*_{\bf k'}(\tau') \right]\;,
\label{xiqxiqf}
\end{equation}
which can also be expressed in terms of the variable $z=k/(aH)$.
Using the definition of the inflaton power spectrum
$P_{\delta\varphi}$ in terms of the two-point correlation function for
the fluctuations as~\cite{liddlebook}

\begin{equation}
P_{\delta\varphi} = \frac{k^3}{2 \pi^2}\int \frac{d^3k'}{(2 \pi)^3}
\langle \delta \varphi ({\bf k},z) \delta \varphi ({\bf
  k}',z)\rangle\;,
\label{spectrum}
\end{equation}
we then obtain for the quantum noise contribution for the power
spectrum the result: 

\begin{eqnarray}
P_{\delta\varphi}^{\rm (qu)} &\equiv& \frac{k^3}{2 \pi^2 H^4}\int
\frac{d^3k'}{(2 \pi)^3} \int_{z}^{\infty} dz_2 \int_{z}^{\infty} dz_1
G(z,z_1)G(z,z_2)(z_1)^{-1-2\nu}(z_2)^{-1-2\nu} \langle
\tilde{\xi_q}({\bf k},z_1) \tilde{\xi_q}({\bf k}',z_2)\rangle
\nonumber\\  &=& \left[2n(k) +1 \right] \frac{k^3}{2 \pi^2} |
F_k(z)|^2 \;,
\label{Pquantum}
\end{eqnarray}
where

\begin{eqnarray}
F_k(z) &=& \int_z^\infty dz' G(z,z') (z')^{1-2\nu}
\left\{\left[W_{,z'z'}(z') - \frac{2+3Q}{z'} W_{,z'}(z') - \frac{3
    \beta Q}{(1+Q)(z')^2}W(z')  \right] \phi_{{\bf k}}(z')
\right. \nonumber \\ &+&  \left. 2 W_{,z'}(z') {\phi_{\bf
    k}}_{,z'}(z')\right\}\;,
\label{Fkz}
\end{eqnarray}
and $G(z,z')$ is given by Eq. (\ref{Gzz}).

By integrating by parts the term with $W_{,z'z'}$ in Eq.~(\ref{Fkz})
using the identity for the derivative of the Bessel functions (where
$Z_\alpha \equiv J_\alpha, Y_\alpha, H_\alpha^{(1)}$),

\begin{equation}
\frac{d Z_\alpha(z)}{dz} = Z_{\alpha-1} (z)- \frac{\alpha}{z}
Z_\alpha(z)\;,
\label{identity2}
\end{equation}
and using Eq. (\ref{gensolQ}), after some algebra, we obtain for
Eq. (\ref{Fkz}) the complete result in terms of an arbitrary filter
function $W(z)$,

\begin{eqnarray}
F_k(z) &=& \frac{\pi^{3/2} z^\nu H}{4 k^{3/2}} \int_z^\infty dz'
(z')^{3/2-\nu} \left[ J_\alpha(z) Y_\alpha(z') - J_\alpha(z')
  Y_\alpha(z) \right] \nonumber \\ &\times&  \left[ \frac{\beta
    Q}{1+Q} W_{,z'}(z') H_\mu^{(1)}(z') + z' W_{,z'}(z')
  H_{\mu-1}^{(1)}(z') \right. \nonumber \\ &-& \left.  \frac{3Q}{z'}
  \left(\frac{\beta Q}{1+Q} + \eta \right) W(z') H_{\mu}^{(1)}(z') -3Q
  W(z') H_{\mu-1}^{(1)}(z') \right]  \nonumber \\  &-&
\!\!\frac{\pi^{3/2} z^\nu H}{4 k^{3/2}} \!\! \int_z^\infty \!\!\! dz'
(z')^{5/2-\nu} \!\! \left[ J_\alpha(z) Y_{\alpha-1}(z') -
  J_{\alpha-1}(z') Y_\alpha(z) \right] W_{,z'}(z') H_\mu^{(1)}(z').
\label{resFkz}
\end{eqnarray}

If we now make use of the window function (\ref{window}),  the
integrals in Eq.~(\ref{resFkz}) can all be performed.  If we also
neglect terms proportional to the slow-roll coefficients and taking
the leading order terms in the resulting expression,  using that
$0<z<\epsilon<1$, we obtain for Eq.~(\ref{resFkz})  the much simpler
result

\begin{eqnarray}
F_k(z) \approx -i \,\frac{z^{3/2-\mu} H}{\sqrt{2} k^{3/2}} \left[ 1  +
  {\cal O}(\epsilon^2) \right]\;.
\label{finalFkz}
\end{eqnarray}
We note the important result that at leading order $F_k(z)$ is
independent on the dissipation coefficient $Q$. Dependence on $Q$ in
Eq. (\ref{finalFkz}) can be verified that only appears at order ${\cal
  O}(\epsilon^3)$ onwards and it involves, multiplicatively, the
slow-roll coefficients, being, therefore, negligible those
contributions.  Consequently, the quantum component of the spectrum
results to be only  very weakly dependent on the dissipation. 

{}Finally, from Eqs. (\ref{Pquantum}) and (\ref{finalFkz}), the
quantum noise contribution for the power spectrum then becomes

\begin{eqnarray}
P_{\delta\varphi}^{\rm (qu)} \simeq  \coth\left(\frac{zH}{2T} \right)
\frac{z^{3-2\mu} H^2}{4 \pi^2} \;,
\label{Pq2}
\end{eqnarray}
which, as conventional, it is evaluated at the moment  of horizon
crossing, $z \equiv k/(aH) = 1$.  Note that the thermal
enhancement factor in the above result was also found by the authors
in \cite{Bhattacharya:2005wn}, although in there it was assumed
already a thermal initial distribution before inflation and that the
thermal background was assumed decoupled from the inflaton dynamics
(i.e., no nonisentropic or warm inflation regime).  Here, we are only
requiring a thermal radiation bath during inflation (as is generically
the case in warm inflation models) and that the inflaton fluctuations
are in thermal equilibrium with the radiation bath at the time where
the relevant scales leave the horizon, as already discussed before
(for a discussion of the different microscopic processes possibly
contributing and leading to this thermal state, see, e.g.,
Ref.~\cite{dissipation}).
 
Using that $3-2\mu \simeq 2 \eta \ll 1$ and taking $T=0$ in
Eq. (\ref{Pq2}), we obtain the standard, approximately
scale-invariant, spectrum result for cold  inflation~\cite{liddlebook}

\begin{equation}
P_{\delta\varphi}^{\rm (qu)} \stackrel{T\to 0}{\longrightarrow}
\frac{H^2}{4 \pi^2}\;.
\label{Pcold}
\end{equation}

\begin{figure}[htb]
 \centerline{
   \psfig{file=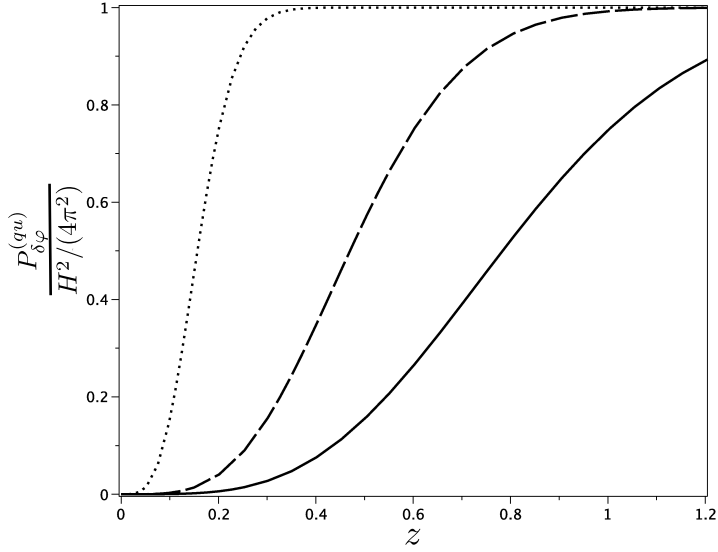,scale=0.50,angle=0} \hspace{0.3cm} }
   \caption{\sf Normalized power spectrum obtained using a Gaussian
     window filter, with $\epsilon=0.5$ (solid line), $0.3$ (dashed
     line) and $0.1$ (dotted line).}
   \label{figfilter}
\end{figure}

If we choose another window function, for example the Gaussian one
Eq. (\ref{gaussian}),  we have to perform the integrals that appear in
Eq. (\ref{resFkz}) numerically.  We have explicitly checked
numerically that for values of the parameter $\epsilon \lesssim 0.1$
in the Gaussian filter  Eq. (\ref{gaussian}), the results obtained for
the power spectrum are in very good agreement with those obtained with
the step function filter, Eq. (\ref{Pq2}).  In fact, when $\epsilon
\lesssim 0.1$, the step filter is a very good approximation to the
Gaussian filter. {}For example, this can be shown from the numerical
results for the normalized spectrum obtained with the Gaussian filter,
with the thermal  contribution factor out, and whose results are
shown in the {}figure \ref{figfilter}. 

\subsection{The thermal noise component for the power spectrum}

Let us now compute the contribution in Eq. (\ref{powercorrelation})
depending on the thermal noise correlation function, given by the
second term on the right-hand-side of
Eq. (\ref{powercorrelation}). {}First note that the thermal noise
correlation function Eq.~(\ref{flucdiss}), in momentum space and in
terms of the variable $z$, can be written as

\begin{equation} \label{xicorrelationz0} 
\langle \tilde{\xi}_T({\bf k},z_1) \tilde{\xi}_T({\bf k}',z_2) \rangle
= 2\Upsilon T  \frac{H^4}{k^2 k'}z_1^3 z_2 \delta\left(z_1-z_2\right)
(2 \pi)^3 \delta({\bf k} + {\bf k}') \;,
\end{equation}
where we have used the time variable written in terms of $z$,

\begin{equation}
t = \ln{\left( \frac{k}{Hz} \right)^{1/H}} \;,
\end{equation}
together with

\begin{equation}
\delta(t_1-t_2) = H \delta \left[ \ln{\left( \frac{kz_2}{k'z_1}
    \right)} \right] \;,
\end{equation}
and the property of the Dirac delta-function, 

\begin{equation}
\delta(f(x)) = \frac{\delta(x-x_0)}{|f'(x)|_{x_0}} \;,
\end{equation}
where $x_0$ are the roots of $f(x)$.

{}From Eq. (\ref{xicorrelationz0}) and using the definition of the
power spectrum Eq. (\ref{spectrum}), we then obtain the contribution
to the power spectrum coming from the thermal noise source as given by

\begin{eqnarray}
P_{\delta\varphi}^{\rm (th)} &\equiv&  \frac{k^3}{2 \pi^2 H^4}   \int
\frac{d^3k'}{(2 \pi)^3} \int_{z}^{\infty} dz_2 \int_{z}^{\infty} dz_1
G(z,z_1)G(z,z_2)\frac{(z_1)^{1-2\nu}}{z_1^2}\frac{(z_2)^{1-2\nu}}{z_2^2}
\langle \xi_T({\bf k},z_1) \xi_T({\bf k}',z_2)\rangle \nonumber \\ &=&
\frac{\Upsilon T}{\pi^2} \int_{z}^{\infty}  dz' z'^{2-4\nu} G(z,z')^2
\;.
\label{Pthermal}
\end{eqnarray}

Moss and Graham in Ref. \cite{GM} have also found a similar result for
the power spectrum, except that they have, instead of $\Upsilon$ in
their result,  an effective dissipation coefficient $\Upsilon_{\rm
  eff}$. They then determine this effective dissipation coefficient by
a matching procedure with the expected spectrum obtained for a flat
space thermal field theory.  Here, as we are going to see below, this
is not necessary, since we are already determining the complete
inflaton power spectrum, that includes both thermal and quantum noise
sources.

{}For the relevant observational scale for the inflaton perturbations,
which are produced in the late stages of inflation, the value of the
argument $z$ in (\ref{Pthermal}) can be taken as been small.  Then,
for example, we can approximate the integral appearing in
Eq.~(\ref{Pthermal}) such that

\begin{eqnarray}
\int_{z}^{\infty} dz' z'^{2-4\nu} G(z,z')^2  &\approx &
\frac{\pi^2}{4} \left[z^{\nu} Y_\alpha(z) \right]^2 \int_0^\infty dz'
(z')^{2-2\nu} J_\alpha^2(z') \nonumber \\ &=& \frac{\pi^2}{4}
\left[z^{\nu} Y_\alpha(z) \right]^2   \frac{\Gamma(\nu-1)
  \Gamma(\alpha-\nu+3/2)}{ 2\sqrt{\pi} \, \Gamma(\nu-1/2)
  \Gamma(\alpha+\nu-1/2)}\;,
\label{intG2approx}
\end{eqnarray}
where $\Gamma(x)$ is the Gamma function. Likewise, we can also use the
asymptotic form for the Bessel function, i.e.,

\begin{equation}
Y_\alpha(z) \simeq -\left(\frac{2}{z}\right)^\alpha
\frac{\Gamma(\alpha)}{\pi}\;.
\label{Yapprox}
\end{equation}

By neglecting terms proportional to the slow-roll coefficients, with
$\eta,\, \beta \ll 1 $, such that we can take $\alpha \simeq \nu=
3(1+Q)/2$ in the arguments of the Gamma functions in
Eqs. (\ref{intG2approx}) and (\ref{Yapprox}), the thermal component
for the power spectrum, Eq. (\ref{Pthermal}), can finally  be
approximated as

\begin{eqnarray}
P_{\delta\varphi}^{\rm (th)}(z) &\simeq &  \frac{\Upsilon T}{16\pi^2}
z^{2 \nu-2 \alpha} \frac{\left[2^{\nu} \Gamma(\nu)\right]^2
  \Gamma(\nu-1)} {\Gamma(2 \nu -1/2) \Gamma(\nu-1/2)}  \nonumber\\ &=&
z^{2 \eta - 2 \beta Q/(1+Q)} \frac{H T}{\pi^2}  \frac{3 Q \, 8^Q \,
  \left[\Gamma(3/2+3Q/2)\right]^3} {(1+3Q)
  \Gamma(5/2+3Q)\Gamma(1+3Q/2)}\;.
\label{Pthermal2}
\end{eqnarray}
Note that the thermal spectrum is still approximately scale-invariant,
as we would expect.


\subsection{The total power spectrum and 
relevant cosmological parameter results from observations}
\label{datasec}

{}From the result for the quantum contribution to the power spectrum
Eq.~(\ref{Pq2})  and the result for the thermal contribution,
Eq.~(\ref{intG2approx}) with  Eq.~(\ref{Yapprox}),  our result for the
total power spectrum becomes,

\begin{equation}
P_{\delta\varphi}(z) = \frac{HT}{4\pi^2}  \left[
  \frac{3Q}{2\sqrt{\pi}} 2^{2\alpha}z^{2\nu -2\alpha}
  \frac{\Gamma\left( \alpha  \right)^2\Gamma\left( \nu -1\right)
    \Gamma\left(\alpha-\nu+3/2\right)}{\Gamma\left(\nu
    -\frac{1}{2}\right) \Gamma\left(\alpha+\nu-1/2\right)} +
  \frac{H}{T}\coth\left(\frac{zH}{2T} \right) z^{2\eta} \right] \;,
\label{Pphi}
\end{equation}
which is conventionally evaluated at the moment  of horizon crossing,
$z \equiv k/(aH) = 1$.  Note that in the zero temperature limit,
$T\to 0$, we also have that~\cite{BMR} $\Upsilon \to 0$ ($Q\to 0$) and
the cold inflation result, Eq.~(\ref{Pcold}) is re-obtained as
expected. In the limit of $Q \ll 1$ and $T \gg H$, we obtain
$P_{\delta\varphi} \propto H T$, which is the original result obtained
for warm inflation by Berera and Fang in Ref.~\cite{firstWI}. In the
limit $Q \gg 1$ and  $T \gg 1$ we obtain that $P_{\delta\varphi}
\propto T \sqrt{H \Upsilon}$, which is the result found by Berera and
Taylor in \cite{warmpert1} and also by Hall, Moss and Berera in
\cite{warmpert2}.

{}From a perturbed Friedmann-Robertson-Walker metric with spatial
curvature perturbation $\psi$, it can be defined a gauge invariant
comoving curvature  perturbation as ${\cal R}({\bf x}, t) =  \psi({\bf
  x},t) + H \delta \varphi({\bf x},t)/\dot \phi$, which remains
conserved throughout the evolution on super horizon scales.  We can
then choose a gauge, for example a spatially flat gauge, and write the
curvature perturbation simple as~\cite{liddlebook} ${\cal R} = H
\delta \varphi/\dot \phi$.  Thus, from Eq.~(\ref{Pphi}), we obtain the
amplitude for the curvature perturbation during inflation,
$\Delta_{\cal R}$, as 

\begin{equation}
\Delta_{\cal R}^2 = \frac{H^2}{\dot \phi^2} P_{\delta\varphi} =
\Delta_{\cal R}^2(k_0) \left(\frac{k}{k_0}\right)^{n_s-1}\;,
\label{powerlaw}
\end{equation}
where $k_0$ is a pivot scale~\cite{liddlebook}, $k_0 \equiv 0.002\,
{\rm Mpc}^{-1}$, $n_s$ is the spectral index, defined by

\begin{equation} \label{fullns}
n_s-1 = \frac{d \ln \Delta_{\cal R}^2}{d N_e} = \frac{d \ln
  \Delta_{\cal R}^2}{d \ln k}\;,
\end{equation}
where $N_e$ denotes the number of e-folds of inflation.  These are all
quantities that are read at the longest length scales of interest and
taken to be those that first crossed the Hubble radius and which
corresponds to the scales that crossed the horizon $N_*$ e-folds
before the end of inflation. $N_*$ is between the usual canonical
values, $N_*=$ 50 and 60 e-folds~\cite{Nespectra}, whose specific
value depends on the details of inflation and on the reheating phase
after inflation.  These will be the values of e-folds before the end
of inflation that we will  consider in the numerical analysis to be
shown in section~\ref{sec5} below.

The most recent CMB radiation measurements from the WMAP  nine years
of accumulated data (WMAP9yr) gives for $\Delta_{\cal R}^2(k_0)$  and
$n_s$, respectively, the values~\cite{WMAP9yr}:  $\Delta_{\cal
  R}^2(k_0) = (2.41 \pm 0.10)\times 10^{-9}$ and $n_s= 0.972 \pm
0.013$, for the measurements from WMAP only. When the results from
WMAP are combined with the results obtained from the extended CMB data
(eCMB) (the CMB data coming from the Atacama Cosmology Telescope (ACT)
and the South Pole Telescope (SPT)), from  baryon acoustic oscillation
(BAO) results and measurements for $H_0$ from galaxies surveys, it is
obtained (assuming also contributions from tensor modes) $\Delta_{\cal
  R}^2(k_0) = (2.464 \pm 0.072)\times 10^{-9}$ and $n_s = 0.9636 \pm
0.0084$, respectively,  which slight improves over the previous WMAP
seven years of accumulated results (WMAP7yr),  combined with similar
databases (WMAP+BAO+$H_0$), which have produced the
results~\cite{WMAP} $\Delta_{\cal R}^2(k_0) = (2.430 \pm 0.091)\times
10^{-9}$ and $n_s=0.968 \pm 0.012$.

Another cosmological parameter that helps to constrain many inflation
type of models is the tensor to scalar curvature perturbation ratio,
$r$, 

\begin{equation}
r \equiv \frac{\Delta_h^2(k_0)}{\Delta_{\cal R}^2(k_0)}\;,
\label{tensor}
\end{equation}
where $\Delta_h^2(k)$ is the power spectrum of tensor metric
perturbations, which for cold inflation is given by~\cite{liddlebook}  

\begin{equation}
\Delta_h^2 = \frac{8}{M_P^2} \frac{H^2}{4 \pi^2}\;,
\label{Deltahcold}
\end{equation}
where $M_P \equiv 1/\sqrt{8 \pi G}$  is the reduced Planck mass.  The
value of $r$ is still poorly known, but an upper limit can be found
from the observational data.  The recent WMAP9yr results
(WMAP+eCMB+BAO+$H_0$) gives $r<0.13$ (at $95\%$ CL).

It has been shown in Ref.~\cite{mohanty} that in a thermal environment
there can be a contribution to the tensor
perturbation~(\ref{Deltahcold}) coming from stimulated emission, if
gravitons were in thermal equilibrium  with the radiation bath. This
effect would then imply a thermal  enhancement factor $\coth\left[z
  H/(2T) \right]$ in Eq.~(\ref{Deltahcold}),  just like the same
factor also appearing in the quantum scalar  perturbations,
Eq.~(\ref{Pq2}).  The presence of this thermal enhancement factor in
the tensor perturbation power spectrum depends whether the
gravitational wave field is in thermal equilibrium with the radiation
bath or not.  In the original assumption made by Bhattacharya, Mohanty
and Nautiyal in Ref.~\cite{mohanty}, inflation was considered to start
in a state that was initially in a radiation dominated epoch, with
temperature $T$. Alternatively, if a thermal radiation bath develops
in the course of inflation (as assumed in warm inflation)  and the
gravitons  have enough time to equilibrate with the radiation bath and
be strongly enough coupled to the particles in the radiation bath,
then we also expect them to develop a thermal radiation-like spectrum,
though in this case we would typically expect to be necessary at least
temperatures for the radiation bath of order~\cite{coles} $T \simeq
M_P$, which may be too high compared to the typical temperature scales
during warm inflation, which are of the order of the Grand Unification
(GUT) scale and below it. Nevertheless, the presence of the thermal
enhancement factor  $\coth\left[z H/(2T) \right]$ appearing
multiplicatively in Eq.~(\ref{Deltahcold}) would increase the tensor
to scalar curvature perturbation ratio and can potentially produce a
much stronger constraint on nonisentropic inflation models than in the
absence of this factor.  Since some authors working on warm inflation
have included explicitly in their studies this factor in the tensor
curvature spectrum (see, for example,
Refs.~\cite{delCampo:2007cy,Setare:2012fg}) and given the status of
the presence of this term is unclear in the context of nonisentropic
inflation in general, we will consider cases with and without the
thermal enhancement factor included in the tensor curvature spectrum
in the results to be presented below in section \ref{sec5}, so both
situations can be compared with the present observational constraint
on $r$.

The specialized expression for $r$ in the context of stochastic
nonisentropic inflation model, with the thermal  enhancement factor
included in Eq.~(\ref{Deltahcold})  (and with the spectra taken at the
moment our current horizon scale crossed the de Sitter horizon  during
inflation, i.e., $z_* \equiv k_*/(a_* H_*) = 1$) is then
given~by 

\begin{equation} \label{rswi}
r = \frac{4\varepsilon H^2}{(1+Q)^2\pi^2 P_{\delta\varphi} }
\coth\left(\frac{z_* H}{2T} \right)\Bigr|_{z_*=1} \;,
\end{equation}
where we have used the expression for the amplitude of scalar
curvature perturbation Eq.~(\ref{powerlaw}), along with the slow-roll
equation for $\dot{\phi}$ coming from Eq.~(\ref{varphislow}) and the
equation defining the slow-roll coefficient $\varepsilon$,
Eq.~(\ref{sra1}).

It is useful to check some of the limiting behaviors for $r$ with
respect to the dissipation and temperature.  The result for $r$
becomes more transparent if we instead look at the inverse ratio
$1/r$. Using that $P_{\delta\varphi}= P_{\delta\varphi}^{\rm (qu)} +
P_{\delta\varphi}^{\rm (th)}$, we obtain (at $z=z_* =1$),

\begin{equation}
\frac{1}{r} = \frac{(1+Q)^2}{16 \varepsilon} +  \frac{(1+Q)^2\pi^2
  P_{\delta\varphi}^{\rm (th)}\Bigr|_{z_*=1}} {4\varepsilon H^2
  \coth\left(\frac{H}{2T}\right) }\;.
\label{1/r}
\end{equation}
Taking for example the weak dissipation regime $Q\ll 1$ and for either
$T \ll H$ or $T \gg H$, we obtain $1/r = 1/(16 \varepsilon)  +{\cal
  O}(Q^2)$, i.e.,  $r \simeq 16 \varepsilon$, which is the standard
cold inflation result~\cite{liddlebook}.  {}For $Q \gg 1$ and $T \gg
H$, we obtain that  $1/r  = \sqrt{3\pi} Q^{5/2} /(32 \epsilon) +{\cal
  O}(Q^2)$, i.e.,  $r \simeq 32 \epsilon/(\sqrt{3\pi}  Q^{5/2})$,
while in the case of not having the thermal enhancement factor in
Eq.~(\ref{rswi}), or also when $T\ll H$ in (\ref{rswi}), we obtain the
warm inflation result for $r$ in the strong dissipation regime, $r
\simeq  16 \epsilon H/(\sqrt{3\pi} T Q^{5/2})$.  We can then conclude
that even in the presence of the thermal enhancement factor in $r$,
for the weak dissipation regime $Q \ll 1$ the cold inflation result
for $r$ is preserved independently of the value of the temperature for
the thermal radiation bath.  It is only in the strong dissipation
regime $Q \gg 1$ that the ratio is affected  by the dissipation, which
makes $r$ smaller by a factor  $Q^{5/2}$.  This leads to an important
aspect of warm inflation, making it possible to evade the so called
Lyth bound~\cite{Lyth} (in particular, as observed in
Ref.~\cite{Cai:2010wt}, this implies that large field inflation models
do not necessarily yield a large tensor-to-scalar ratio).  Therefore,
thermal noises only affect scalar type curvature perturbations,  but
primordial gravitational waves (and associated tensor perturbations)
are still originated from quantum vacuum fluctuations (though it may
receive a correction dependent on the temperature, when in a thermal
bath, because now the vacuum becomes a quantum statistical  vacuum at
some temperature $T$ and the gravitons can have a non zero  occupation
number, as shown in Ref.~\cite{mohanty}).

{}For the sake of completeness, we also show here the limiting
behaviors of $n_s$ in relation to $Q$ and $T$. Details concerning the
derivation of $n_s$, in the context of nonisentropic inflation in
general,  are given in the appendix A.  {}From Eq.~(\ref{fullns}),
taking the $Q=0$ and $T=0$ limit, we obtain the expected result from
cold inflation~\cite{liddlebook},

\begin{equation} \label{ns_Q0step}
n_s = 1 + 2\eta - 6\varepsilon \;.
\end{equation}
In the large $Q$ and $T$ limit (and temperature independent
dissipation),  we recover the standard warm inflation result (e.g.,
from Hall, Moss and Berera Ref.~\cite{warmpert2}):

\begin{equation} \label{ns_Qlargestep}
n_s = 1 + \frac{1}{Q}\left[-\frac{9}{4}\varepsilon -  \frac{9}{4}\beta
  + \frac{3}{2}\eta    \right] + \mathcal{O}(1/Q^{1.5}) +
\mathcal{O}(H/T)  \;.
\end{equation}
{}Finally, in the low dissipation $Q \ll 1$ limit, we obtain the first
order correction  for $n_s$:

\begin{equation} \label{ns_Qll1}
n_s = 1 + 2\eta - 6\varepsilon + (8\varepsilon - 2\eta)Q +
\mathcal{O}(Q^2) \;.
\end{equation}

If we relax on the assumption of a power spectrum that is a pure
power-law, like the one given in Eq. (\ref{powerlaw}), and one admits
a running spectral index, then we can write~\cite{Kosowsky:1995aa} 

\begin{equation}
\Delta_{\cal R}^2(k) =  \Delta_{\cal R}^2(k_0)
\left(\frac{k}{k_0}\right)^{n_s(k_0)-1+\frac{1}{2} \ln(k/k_0) d n_s/d
  \ln k}\;.
\label{running}
\end{equation}
With the possibility of a running, the spectral index tends to be blue
tilted, as opposite to a red tilt ($n_s<1$) in the case of no running.
Some inflation models, like hybrid inflation, in fact prefer a blue
tilted spectral index, while others, like chaotic inflation type of
models, tend to be more red tilted.  Observational data, however, are
still not precise enough to give a definitive answer whether a running
is favorable or not, with no statistically significant  deviation from
a pure power-law spectrum. {}For example, the WMAP9yr results for the
running are only with less than $3\sigma$ significance~\cite{WMAP9yr}.
Tensor mode amplitude and scalar running index parameter, when fitted
together, give~\cite{WMAP9yr}: $r<0.47$ (at 95$\%$ CL),  $d n_s/d \ln
k=-0.040 \pm 0.016$ and $n_s = 1.075 \pm 0.046$ (using the combined
data from WMAP9yr+eCMB+BAO+$H_0$).  {}For the analysis and results to
be presented in the next section, we have chosen to concentrate on the
cases of no running of the spectrum, when  contrasting the results
obtained for nonisentropic inflation with the observational data.


\section{Results and constraints for nonisentropic inflation models from
the observational data}
\label{sec5}

Since the early releases of the WMAP CMBR results, a significant
pressure has been put on inflation models.  Both the spectral index
and the tensor to scalar perturbation ratio results  have shown to be
able to exclude large classes of inflationary models. {}For example,
hybrid inflation models are disfavored for predicting in general a too
blue tilted spectrum, while simple chaotic polynomial type of
potentials,  in the case of cold inflation, even though they have a
preference for a red tilted  spectrum, they have been seen to poorly
fit the observational data (though nonminimal coupling to gravity can
make them compatible with the observational data~\cite{WMAP9yr}).   In
particular, the recent WMAP9yr results already exclude  (at the 68$\%$
CL) all simple chaotic cold inflation polynomial potentials $V(\phi)
\propto \phi^p$, with power $p \geq 1$ (see, for instance,  {}figure~7
in Ref.~\cite{WMAP9yr}). 

Since most of the present constraints from the observational data have
been applied most to cold inflation models, it is also important to
have a throughout analysis of how general nonisentropic models
withstand the recent observational data as well.  Even though some
previous works  (see, e.g.,
Refs. \cite{WIreviews2,delCampo:2007cy,delCampo:2010by}) have
addressed the consistency of warm inflation with the observational
data, only a few examples with very specific values of parameters were
studied. A complete analysis covering larger regions of parameter
space and ranging from cold to warm inflation is still missing.  Since
the larger pressure from the observational date are on chaotic
polynomial type of inflation models, which by themselves represent an
important class of inflaton potential models from a microscopic  model
building perspective, these are, therefore, the type of inflation
models we will be interested in studying next.

In the following we will study nonisentropic chaotic inflation
potential models with a generic potential of the form 

\begin{equation} \label{potential}
V(\phi) = \frac{\lambda M_P^4}{p} \left(\frac{\phi}{M_P}\right)^p\;, 
\end{equation}
where $\lambda$ is a dimensionless parameter fixing the energy scale
and we will consider the cases $p=6$, $4$ and $2$, which are the most
common  type of chaotic inflation potentials used, for example, in
cold inflation studies in general. 

Having fixed the class of inflaton potentials that we will analyze, we
next need also to specify the dissipation coefficient $\Upsilon$
appearing in the expressions.  Nonisentropic inflation can be regarded
as a two-fluid system when evaluating  density perturbations, where
both the inflaton and the radiation fluctuations are effectively
coupled through the temperature dependence of the dissipation
coefficient $\Upsilon$.  It has then  been shown in Ref.~\cite{GM}
that because of this dependence of the dissipation coefficient on the
temperature, the radiation perturbations can back react on the
inflaton perturbations, generating undesirable growth modes. However,
recently in Ref.~\cite{MBR},  it was found that intrinsic microscopic
decay processes in the radiation bath itself can cause  it to slight
depart from equilibrium, giving rise to viscous dissipative effects in
the radiation fluid itself. These viscous effects were then shown to
be able to remove the growing modes and to lead  to a power spectrum
for the perturbations consistent with cases where the temperature was
absent in $\Upsilon$. In this case, where the temperature can be
neglected or it is absent in the expression for the dissipation
coefficient, there are two representative forms of dissipation
coefficient that can emerge and that are consistent with microscopic
derivations (see, e.g., Refs. \cite{HR,Ramos:2001zw}): a constant
dissipation coefficient $\Upsilon\equiv \Upsilon_0$ and a dissipation
that has a quadratic dependence on the inflaton field, $\Upsilon
\propto \phi^2$ (these two forms of dissipation coefficients were also
considered by the authors of the
Refs. \cite{delCampo:2007cy,delCampo:2010by} in their studies
involving warm inflation).  These are, therefore, the two forms of
dissipation coefficients we will use in our analysis, a constant
dissipation coefficient, $\Upsilon = \Upsilon_0$  and also one
proportional to the square of the inflaton amplitude, $\Upsilon =
C_\phi \phi^2/M_P$, with constant coefficient~$C_\phi$.


\subsection{Cold and warm inflation dominated regions}

As a preliminary study, let us first establish which regions of
parameter space, in the plane defined by the dissipation coefficient
over the Hubble parameter,  $Q\equiv \Upsilon/(3H)$, and by the
temperature over the Hubble parameter, $T/H$, the spectrum is
dominated by quantum fluctuations or by  thermal fluctuations.  It has
generically been accepted that warm inflation is determined by the
parameters for which $T > H$, considered the condition for which the
thermal  fluctuations dominate the spectrum. Warm inflation has also
been considered in both the weak dissipation regime, $Q \ll 1$ and in
the strong dissipation regime, $Q \gg 1$.  However, it has never been
exactly established for which values of parameters  we could really
discriminate a cold inflation like spectrum, where quantum
fluctuations dominate over the thermal ones, and the opposite regime
of warm inflation. 

\begin{figure}[th]
\centerline{  \includegraphics[height=.35\textheight]{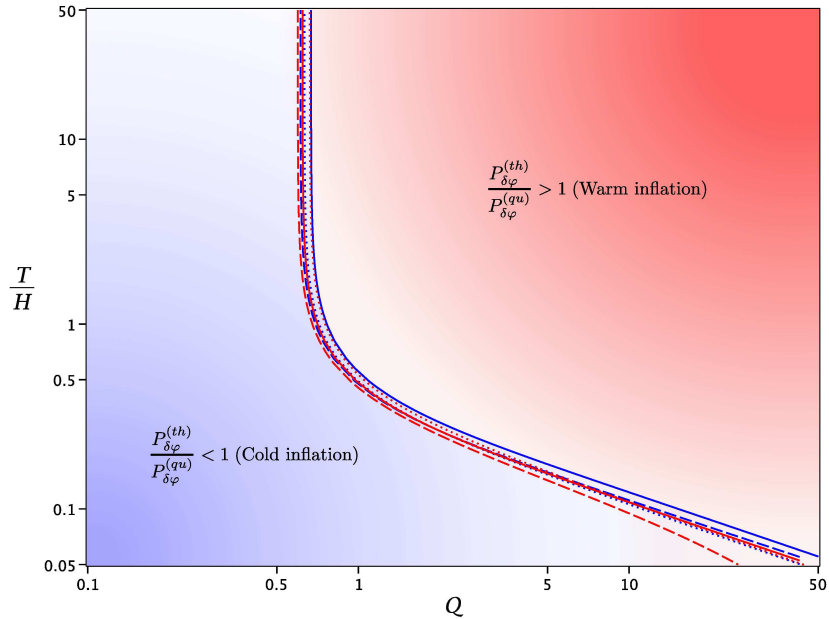}
}
  \caption{Parameter space, in terms of the temperature ratio $T/H$
    and dissipation ratio $Q=\Upsilon/(3 H)$, for quantum fluctuation
    dominated spectrum (cold inflation)  and for a thermal fluctuation
    dominated spectrum (warm inflation).  Red lines are for a constant
    dissipation coefficient, while blue lines are for a  dissipation
    coefficient dependent on the square of the inflaton amplitude.
    The dashed lines are for $p=2$, solid lines are for $p=4$ and
    dotted lines are for $p=6$ (a quadratic, quartic and sextic
    inflaton potentials, respectively).  }
\label{PqoverPt}
\end{figure}

We have determined in the previous section the contributions to the
power spectrum coming from either quantum or thermal sources, with
relevant results given by Eqs. (\ref{Pq2}) and (\ref{Pthermal2}),
respectively. We can, thus, now determine the region of parameters
where one contribution to the spectrum dominates over the other.  In
{}figure \ref{PqoverPt} we show the region of values of $Q$ and $T/H$
for which $P_{\delta\varphi}^{\rm (qu)} > P_{\delta\varphi}^{\rm
  (th)}$, i.e., for which quantum fluctuations dominate over the
thermal fluctuations, and  for the opposite case,
$P_{\delta\varphi}^{\rm (qu)} < P_{\delta\varphi}^{\rm (th)}$,  i.e.,
for which thermal fluctuations dominate over the quantum fluctuations.

We believe that {}figure \ref{PqoverPt} is very illustrative of the
conditions  determining the different regimes that are possible in
nonisentropic inflation models and it represents one of the main
results in this paper.  {}Figure \ref{PqoverPt} shows that there are
no significant differences as regarding the dominating regions for
each of the different polynomial forms for the inflaton potential,
with little differences only between the cases $p=2,\,4,\,6$.  The
figure also clearly shows that it is not always true that whenever $T>
H$ we are in a warm inflation regime of a thermal fluctuation
dominated spectrum. The thermal bath can have a temperature much
larger than the Hubble parameter and we still be in a quantum
fluctuation regime when $Q \lesssim 0.5$. Likewise, the thermal bath
can have in general a very low temperature compared to the Hubble
parameter (or a temperature  less than the one corresponding to the
Hawking temperature) and the system can be in a warm inflation regime,
where thermal fluctuations give the dominant contribution to the
spectrum. These results come crucially from the fact that the quantum
spectrum can depend on the temperature through a thermal  enhancement
factor, as shown in Eq. (\ref{Pq2}), which, as we have already
discussed previously, comes from the consideration that in
nonisentropic inflation models there is the presence of a radiation
bath and that the inflaton fluctuations can be in thermal equilibrium
with this produced radiation bath.
  

\subsection{Observational constraints to the power spectrum}

\begin{figure}[htb]
 \centerline{
   \psfig{file=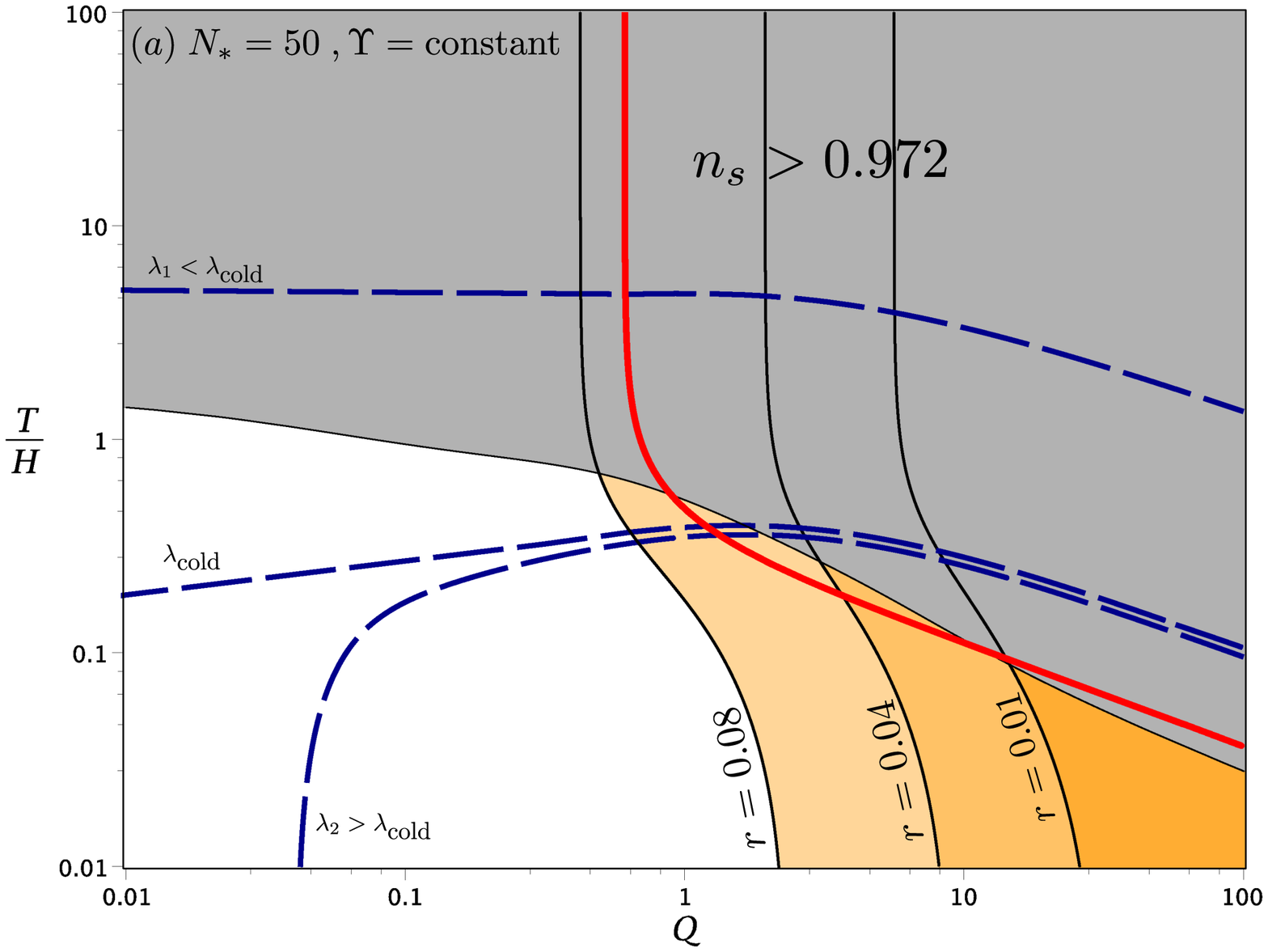,scale=0.4,angle=0}
   \psfig{file=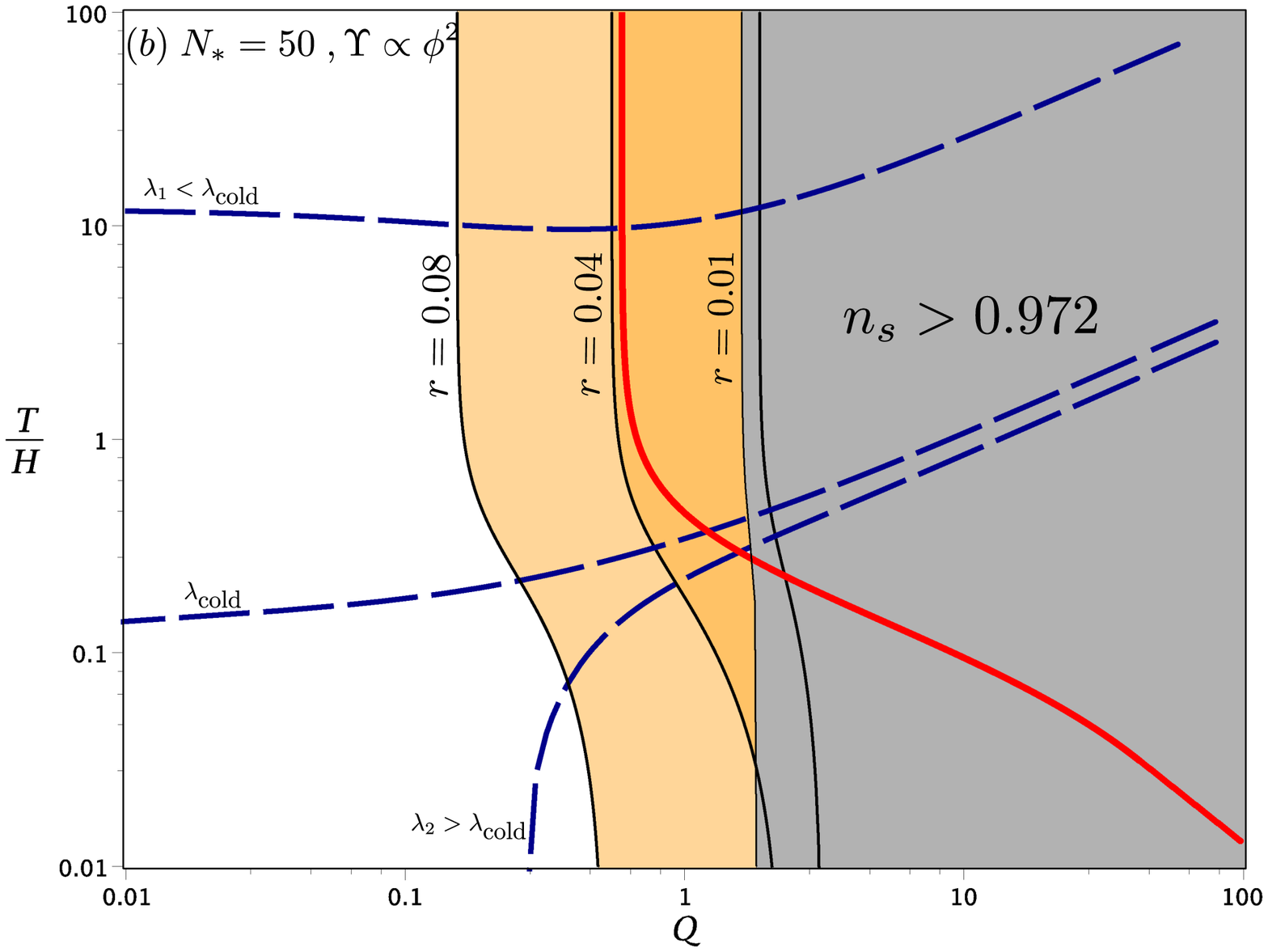,scale=0.4,angle=0}}
\vspace{0.5 cm} \centerline{
  \psfig{file=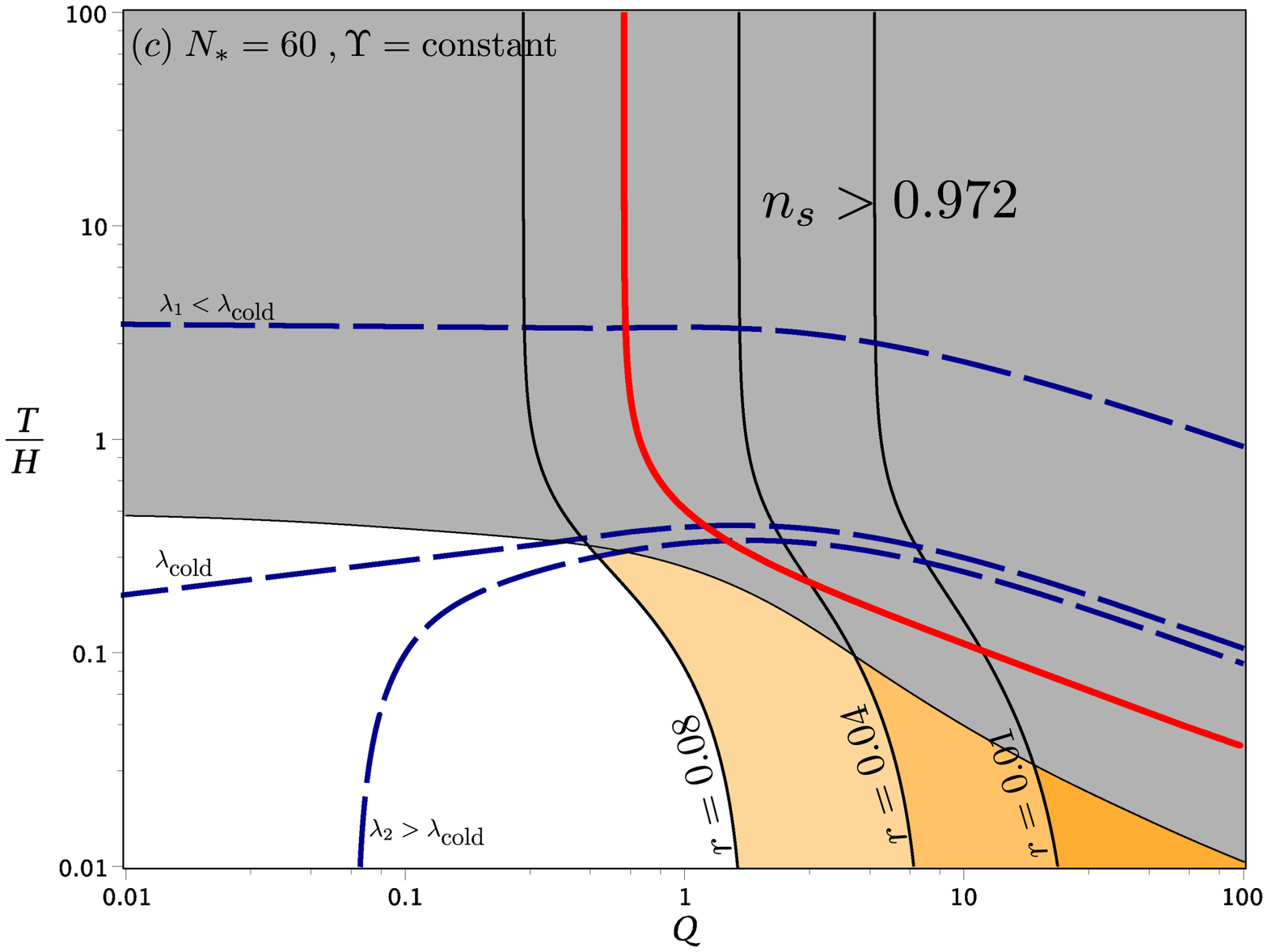,scale=0.4,angle=0}
  \psfig{file=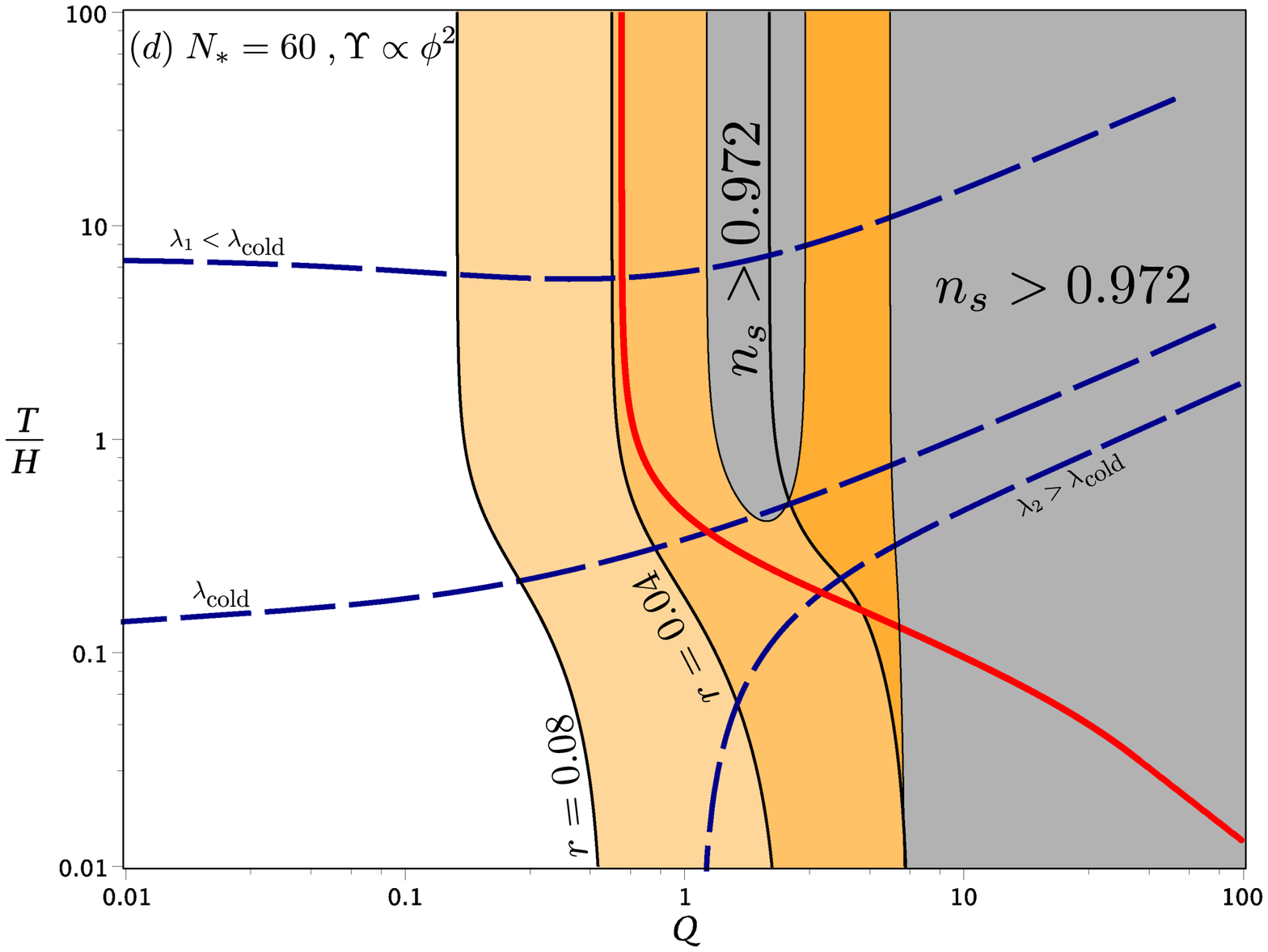,scale=0.4,angle=0} }
\caption{\sf The results for the $V\propto \phi^2$ potential: upper
  panels ((a) and (b)) have horizon exiting fixed at $N_{*}=50$, lower
  panels ((c) and (d)), at $N_{*}=60$. In the panels in the first
  column ((a) and (c)), the dissipation term $\Upsilon$ is kept
  constant, while in the second column the dissipation is
  quadratically dependent on the inflaton field.  In panels (a) and
  (b), $\lambda_1 = 5.0 \times 10^{-12}$, $\lambda_2 = 5.5 \times
  10^{-11}$ and   $\lambda_{\rm{cold}} = 5.13 \times 10^{-11}$. In
  panels (c) and (d),  $\lambda_1 = 5.0 \times 10^{-12}$, $\lambda_2 =
  4.0 \times 10^{-11}$ and   $\lambda_{\rm{cold}} = 3.57 \times
  10^{-11}$. The favorable regions by the observational data are the
  orange ones.}
\label{phi2}
\end{figure}

\begin{figure}[htb]
 \centerline{
   \psfig{file=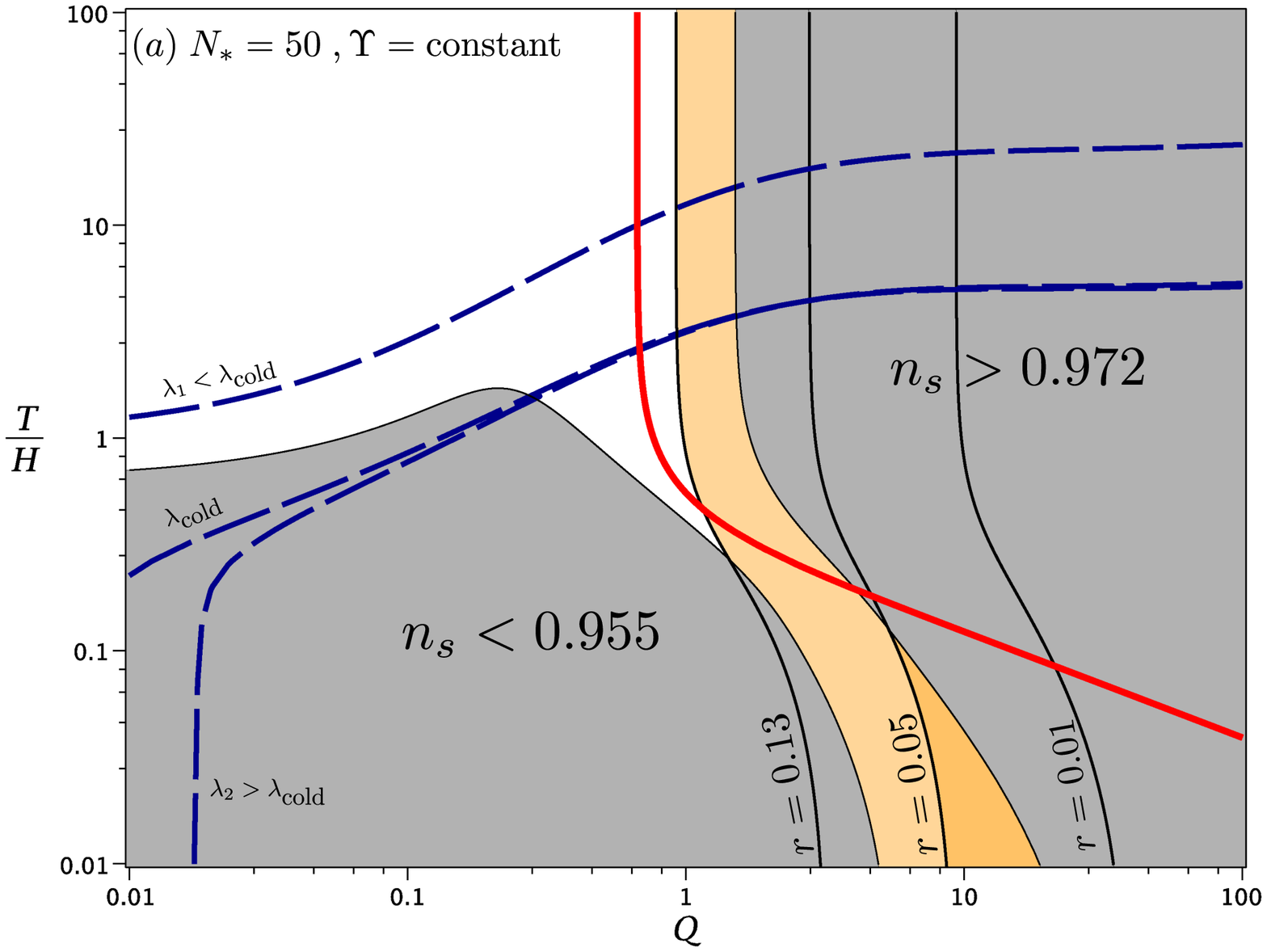,scale=0.4,angle=0}
   \psfig{file=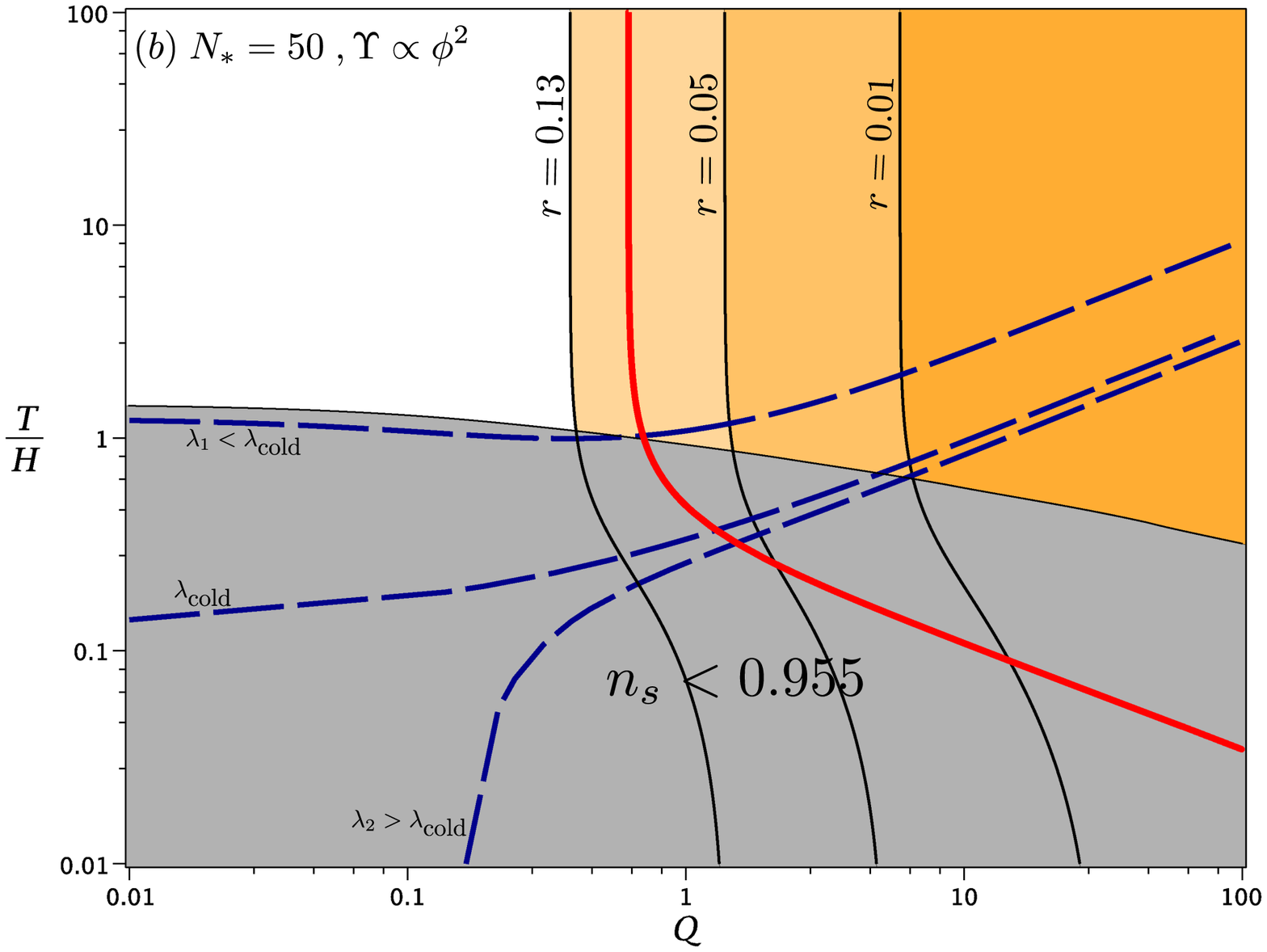,scale=0.4,angle=0}}
\vspace{0.5 cm} \centerline{
  \psfig{file=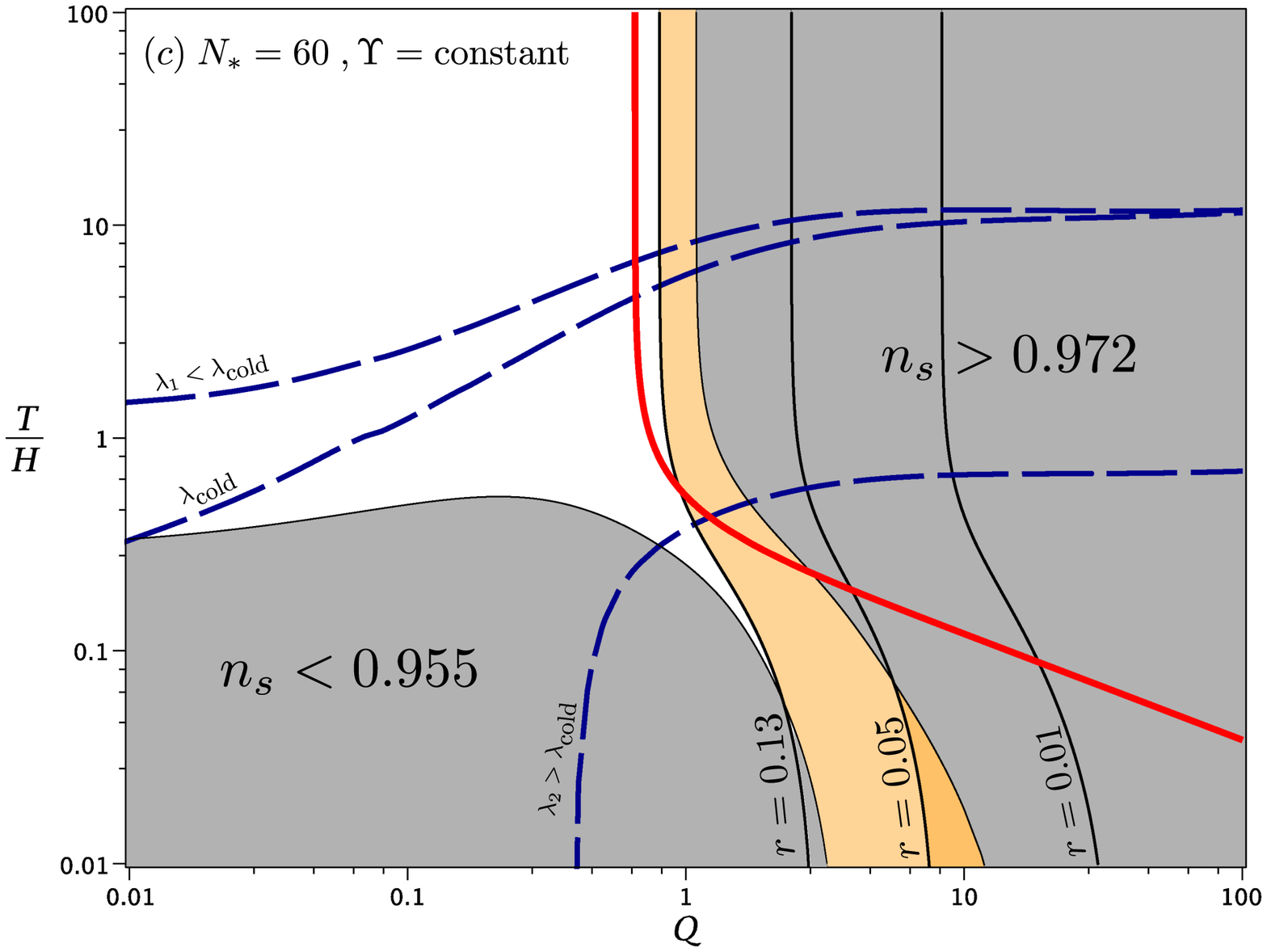,scale=0.4,angle=0}
  \psfig{file=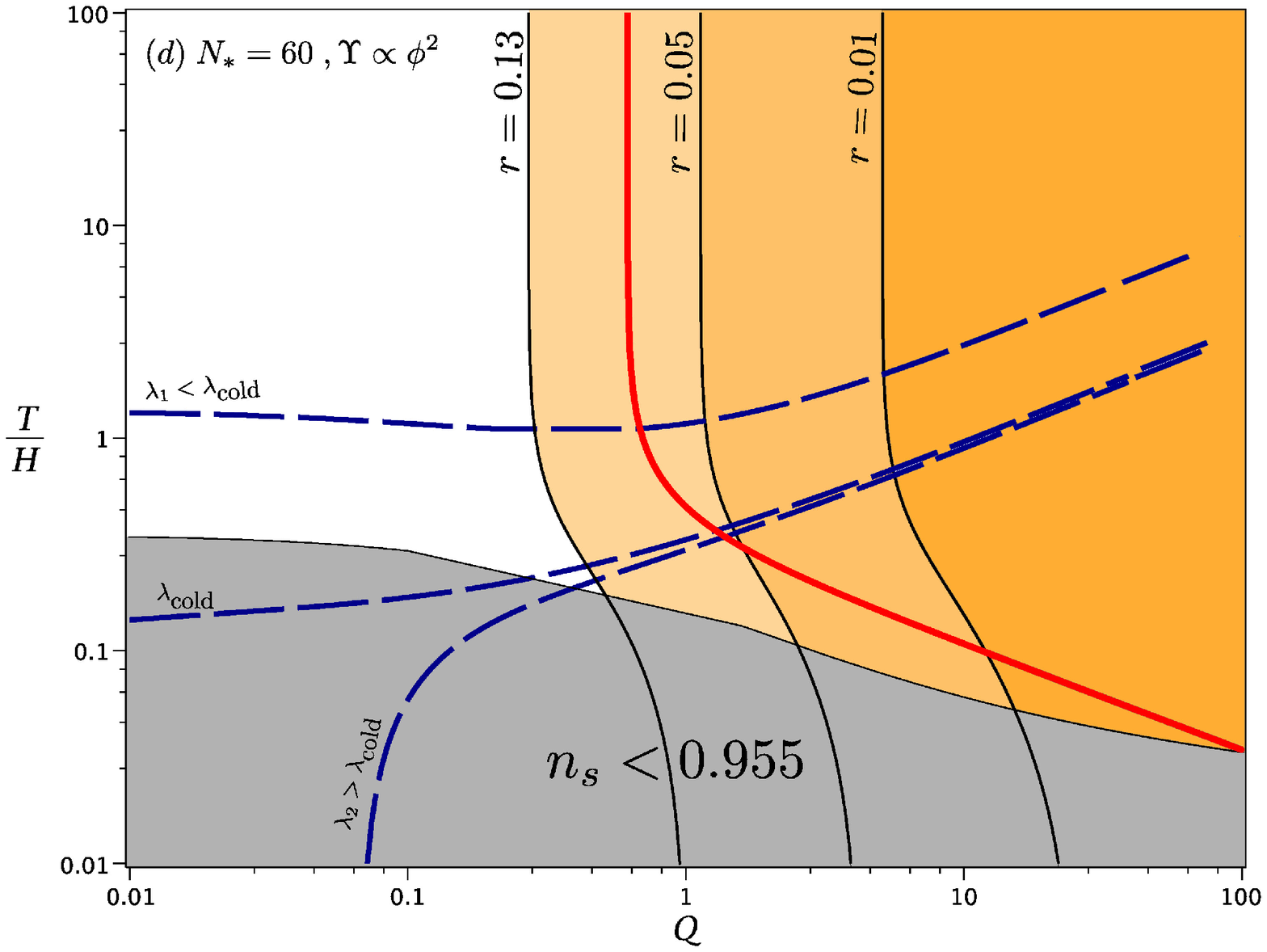,scale=0.4,angle=0} }
\caption{\sf The results for the $V\propto \phi^4$ potential: upper
  panels ((a) and (b)) have horizon exiting fixed at $N_{*}=50$
  e-folds, lower panels ((c) and (d)), at $N_{*}=60$. In the panels in
  the first column ((a) and (c)), the dissipation term $\Upsilon$ is
  kept constant, while in the second column the dissipation is
  quadratically dependent on the inflaton field.  In panels (a) and
  (b), $\lambda_1 =5.0 \times 10^{-14}$, $\lambda_2 =  10^{-12}$ e
  $\lambda_{\rm{cold}} =2.39 \times 10^{-13}$. In panels (c) and (d),
  $\lambda_1 = 5.0 \times 10^{-14}$, $\lambda_2 = 1.5 \times 10^{-13}$
  and   $\lambda_{\rm{cold}} = 1.4 \times 10^{-13}$.The favorable
  regions by the observational data are the  orange ones.} 
\label{phi4}
\end{figure}

\begin{figure}[htb]
 \centerline{
   \psfig{file=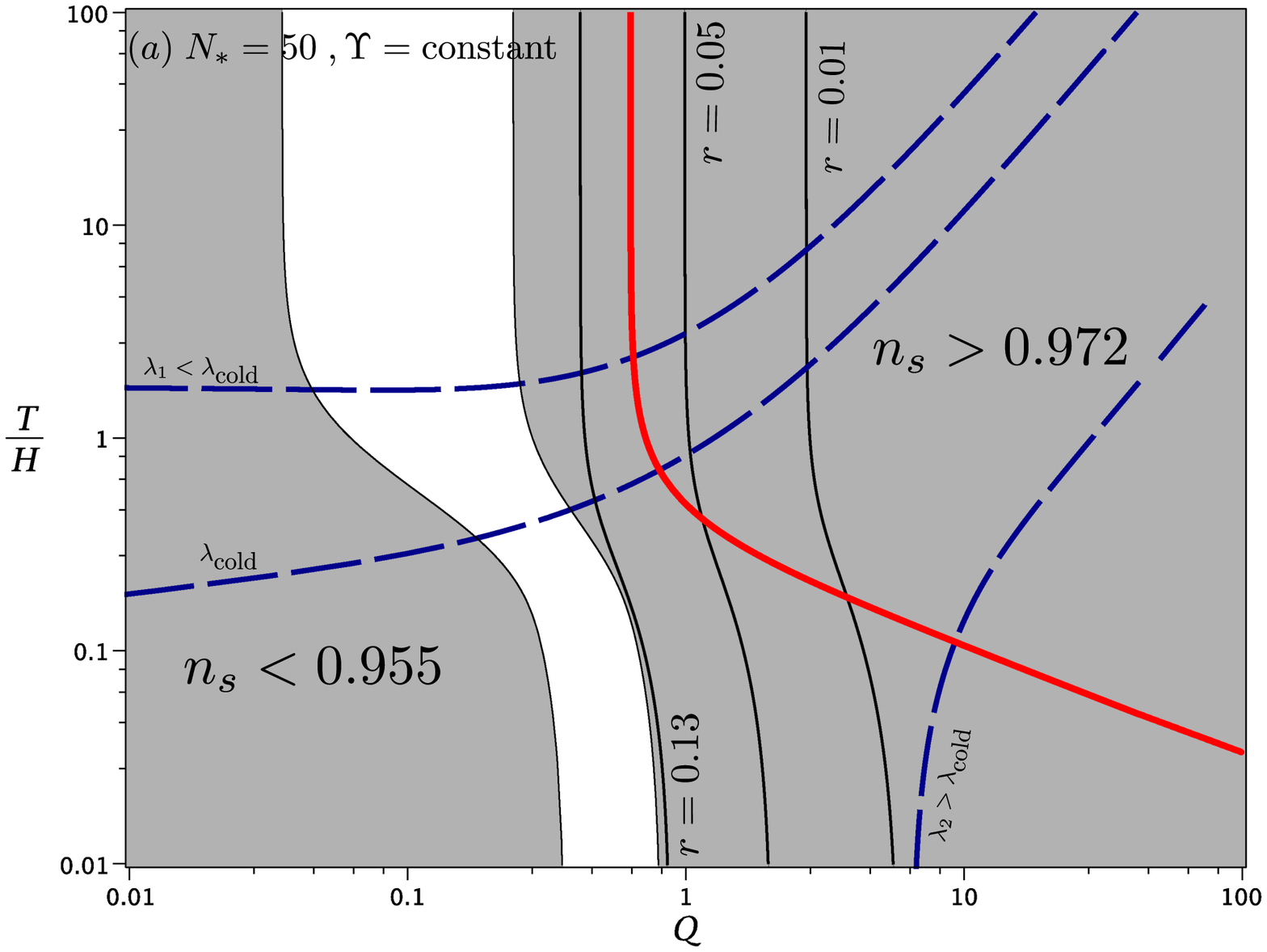,scale=0.4,angle=0}
   \psfig{file=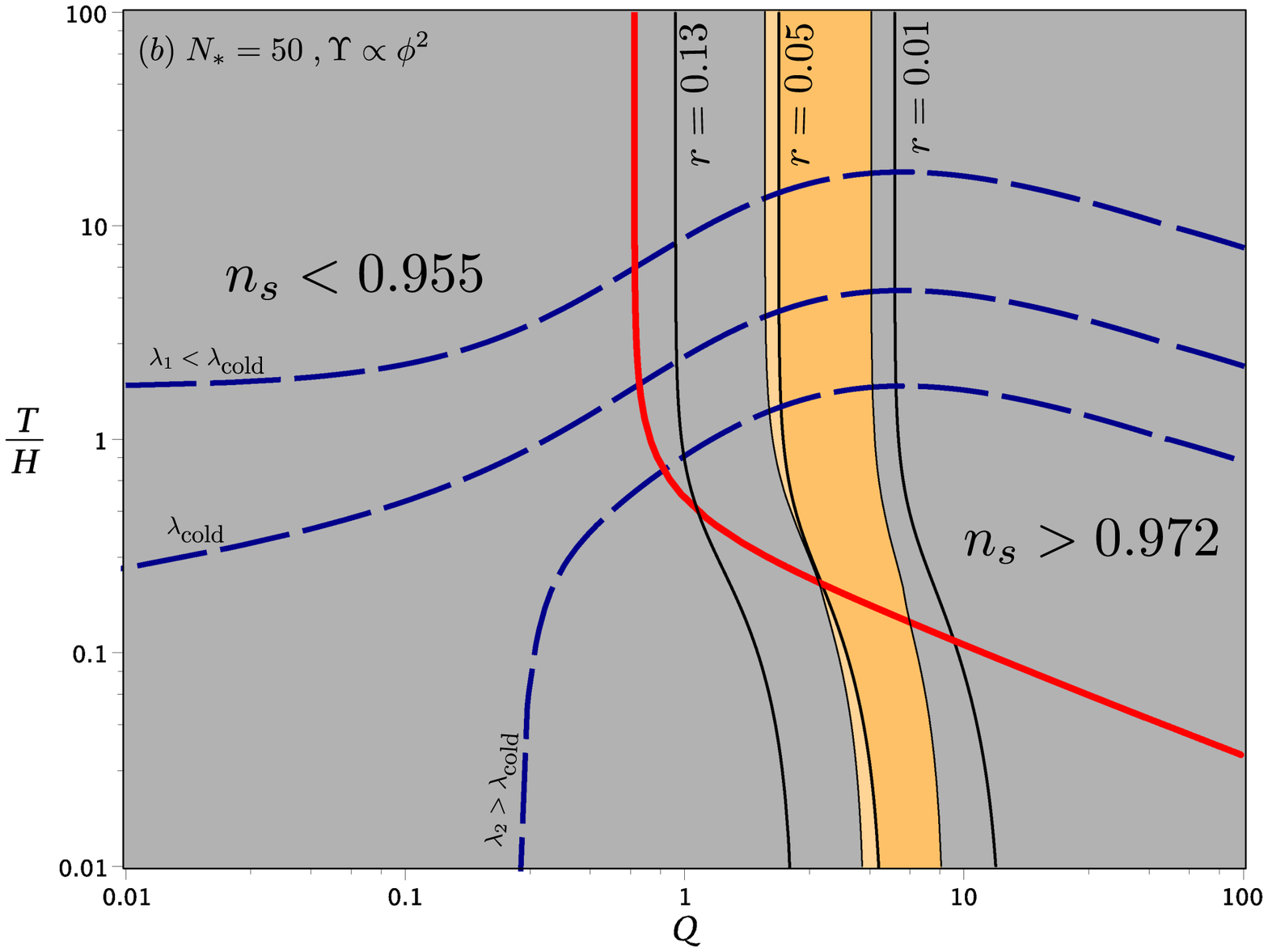,scale=0.4,angle=0}}
\vspace{0.5 cm} \centerline{
  \psfig{file=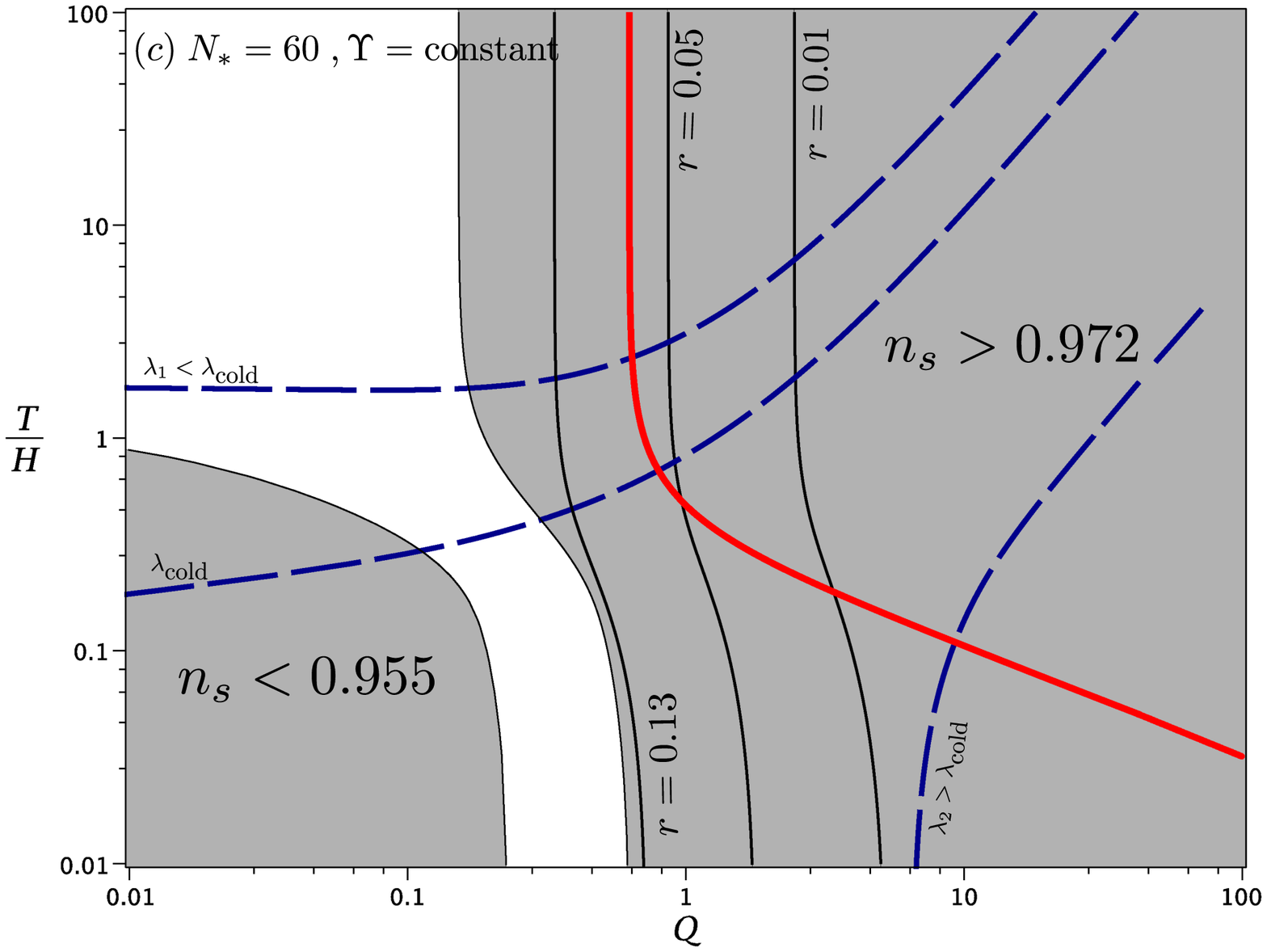,scale=0.4,angle=0}
  \psfig{file=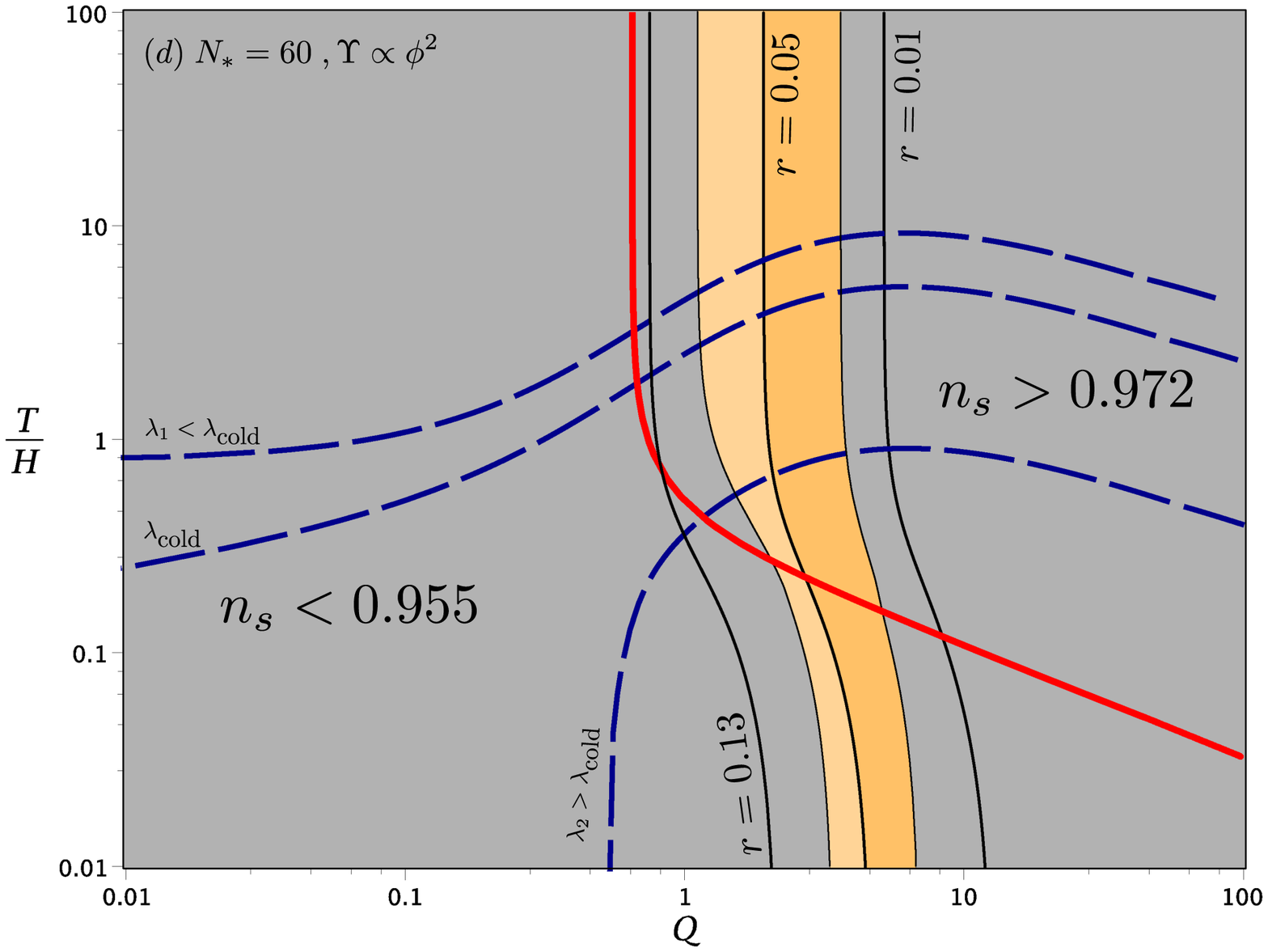,scale=0.4,angle=0} }
\caption{\sf  The results for the $V\propto \phi^6$ potential: upper
  panels ((a) and (b)) have horizon exiting fixed at $N_{*}=50$
  e-folds, lower panels ((c) and (d)), at $N_{*}=60$. First column
  ((a) and (c)), the dissipation term $\Upsilon$ is kept constant,
  while in second column the dissipation is quadratically dependent on
  the field $\phi$.  In panels (a) and (b), $\lambda_1 = 10^{-16}$,
  $\lambda_2 =  10^{-15}$ e   $\lambda_{\rm{cold}} =3.59 \times
  10^{-16}$. In panels (c) and (d),  $\lambda_1 =  10^{-16}$,
  $\lambda_2 =  10^{-15}$ and $\lambda_{\rm{cold}} = 1.78 \times
  10^{-16}$.  The favorable regions by the observational data are the
  orange ones.}
  \label{phi6}
\end{figure}

Having established in the previous subsection which are exactly the
regions of parameter space where quantum or thermal fluctuations
dominate, we turn now our attention to the question of the consistency
of nonisentropic inflation in face of the recent observational data as
coming from WMAP9yr.  The observational parameters we test are the
amplitude of the scalar curvature spectrum  $\Delta_{\cal R}(k_0)$,
the spectral index $n_s$ and the tensor to scalar curvature spectrum
ratio $r$, whose relevant expressions and data we have already given
in section \ref{datasec}.
 
{}For the different forms of polynomial inflaton potentials
considered, with powers $p=2$ (quadratic), $p=4$ (quartic) and $p=6$
(sextic) that we have tested, we have produced plots for the two
canonical values of number of e-folds before the end of inflation,
$N_* =50$ and $N_* =60$. We have read (by numerically solving for) the
spectrum of fluctuations leaving the horizon at  these two values of
$N_*$.  We have also produced sets of plots for the two cases of
dissipation: a constant dissipation coefficient, $\Upsilon\equiv
\Upsilon_0$, and for a dissipation with a quadratic dependence on the
inflaton field, $\Upsilon\equiv C_\phi \phi^2/M_P$, with the constant
coefficients $\Upsilon_0$ and $C_\phi$ fixed by the numerical values
taken for the ratio $\Upsilon/(3H)\equiv Q$.  

\begin{figure}[htb]
 \centerline{
   \psfig{file=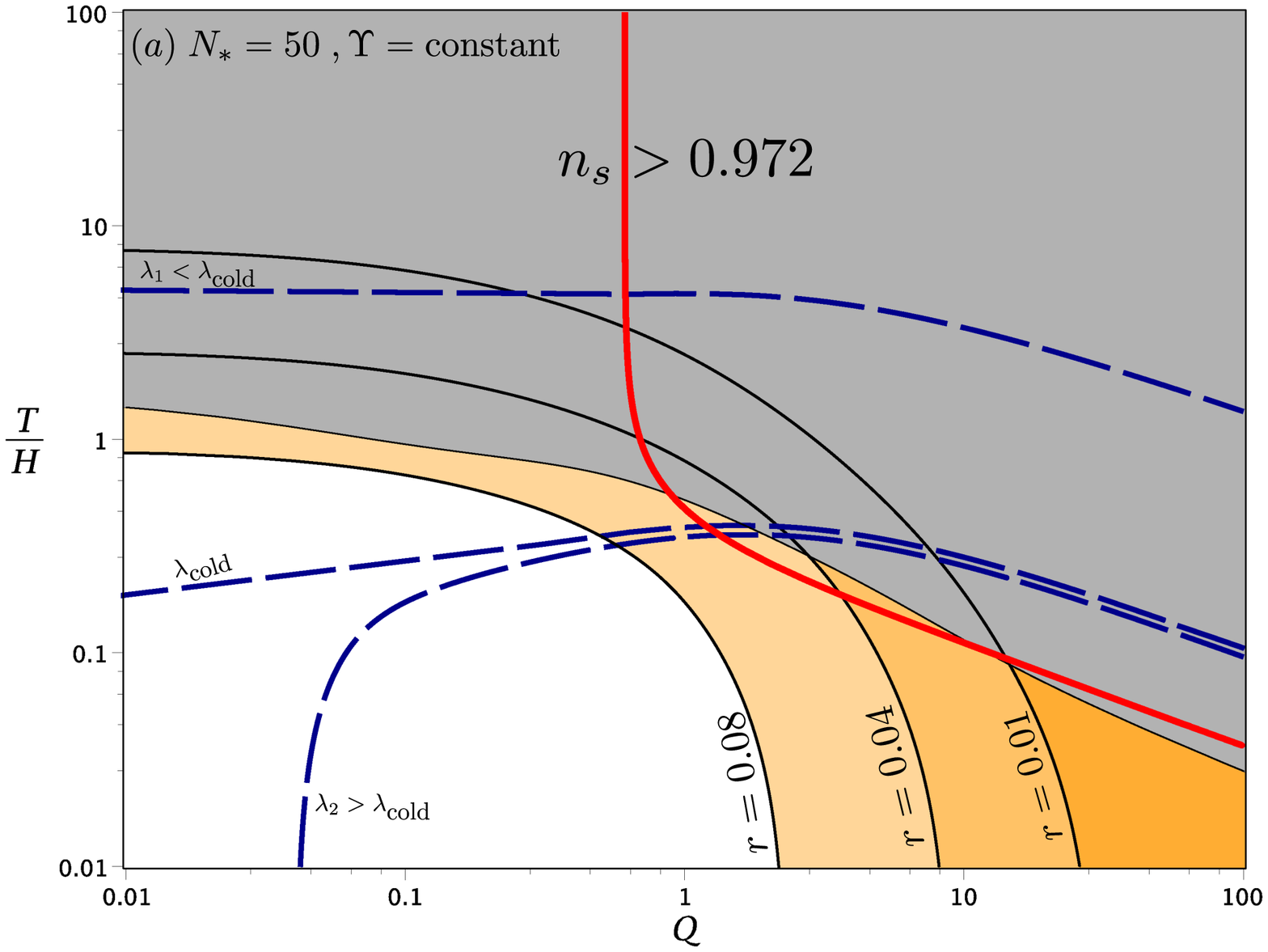,scale=0.4,angle=0}
   \psfig{file=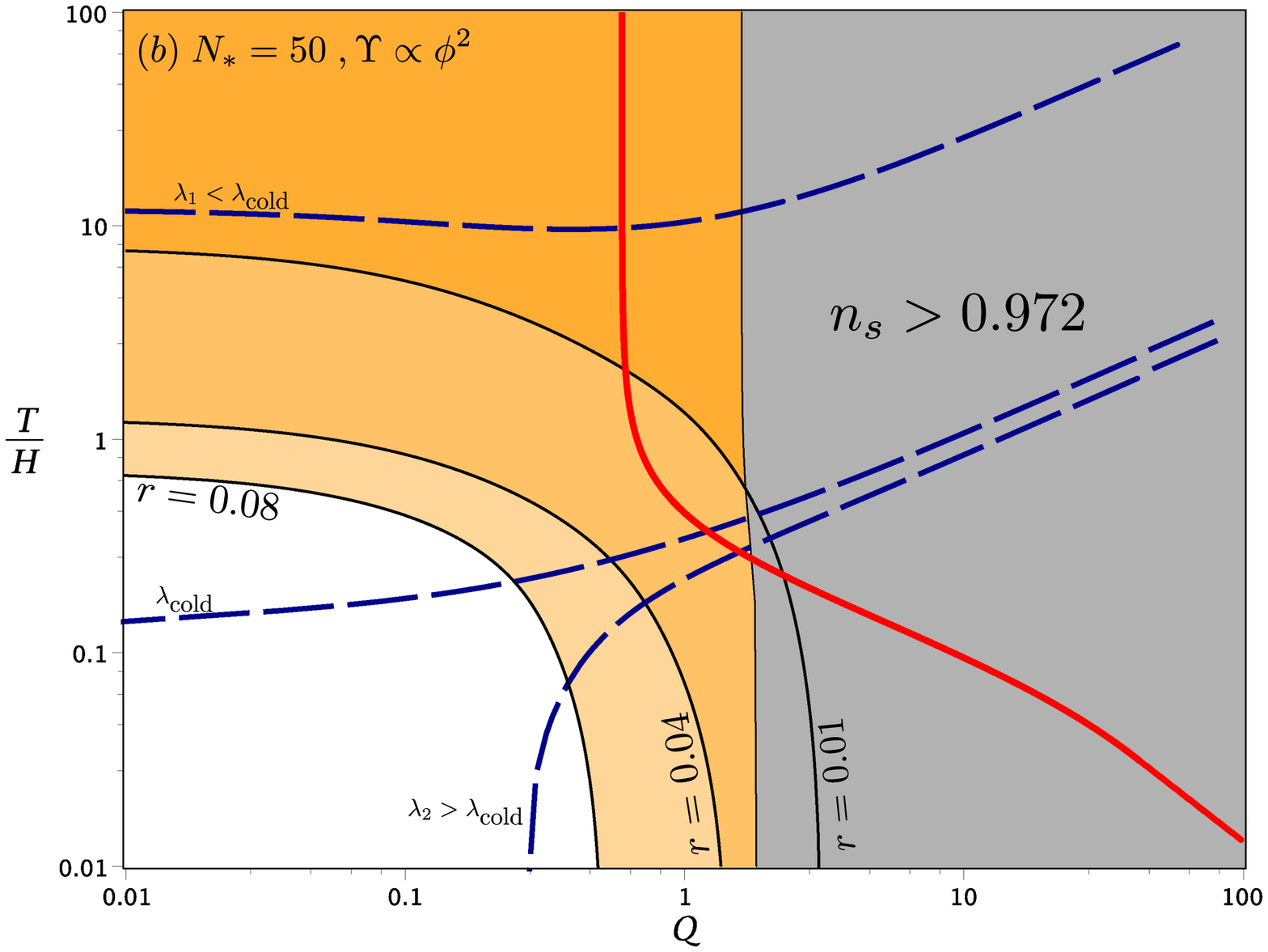,scale=0.4,angle=0}}
\vspace{0.5 cm} \centerline{
  \psfig{file=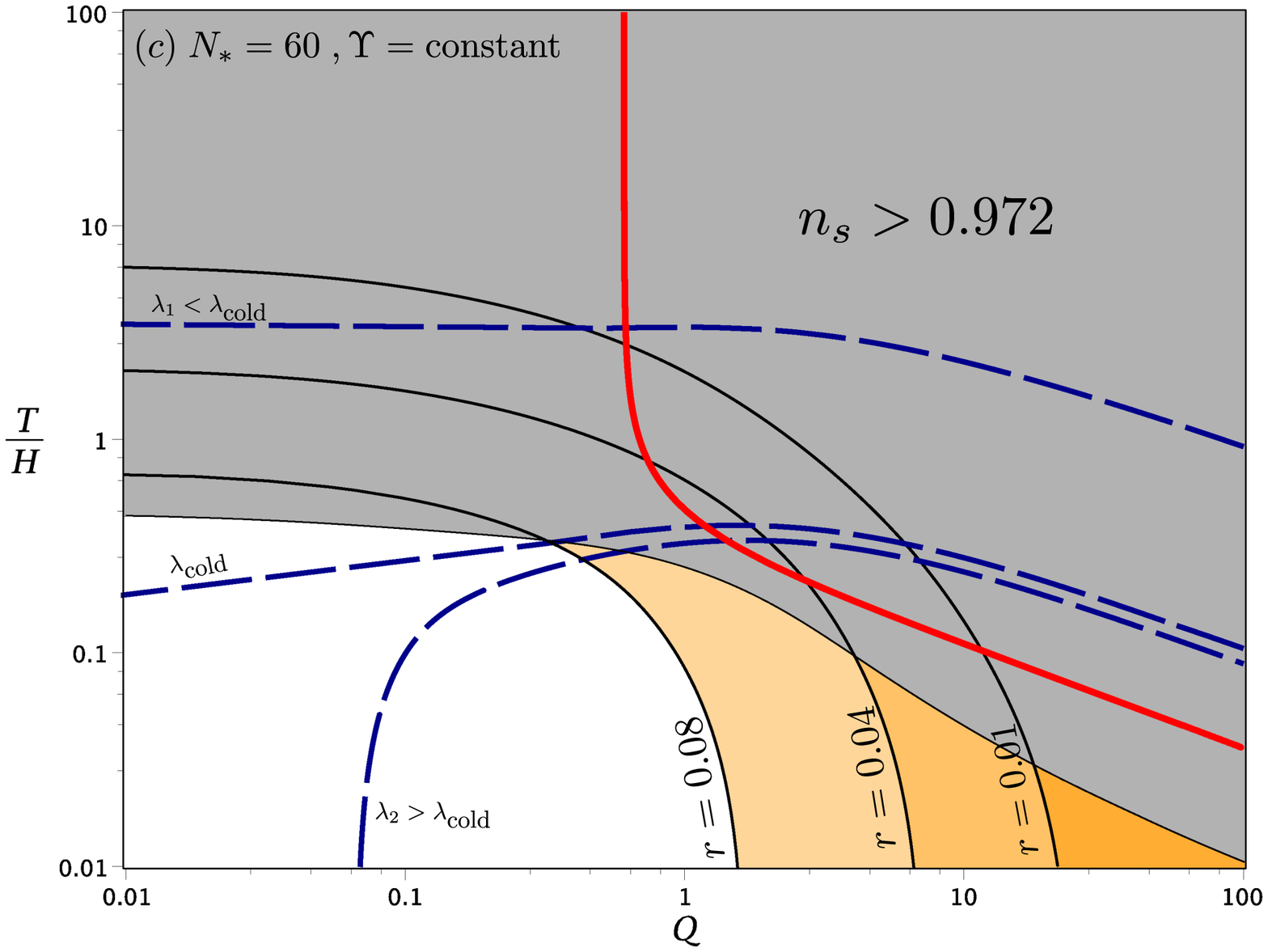,scale=0.4,angle=0}
  \psfig{file=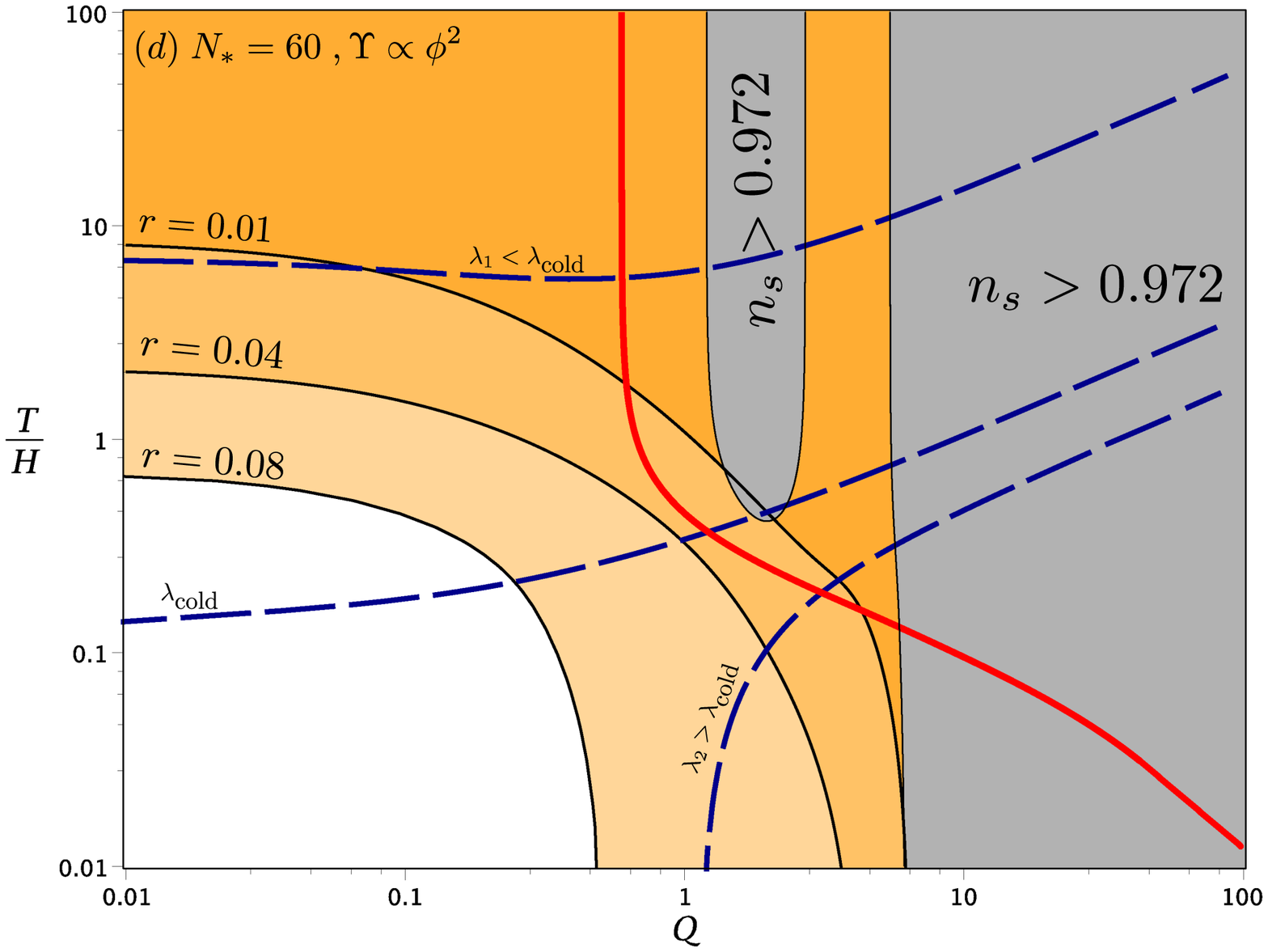,scale=0.4,angle=0}
}
\caption{\sf Results for the $\phi^2$ potential without the thermal
  enhancement factor in the tensor to scalar ratio $r$ expression,
  otherwise, same notation and parameters used in {}figure \ref{phi2}.
  We can observe a much less restriction due to the tensor-to-scalar
  ratio ($r$) and in this case the range of $Q$ is extended to small
  values.  The favorable regions by the observational data are the
  orange ones.}
   \label{phi2noenhance}
\end{figure}

\begin{figure}[htb]
 \centerline{
   \psfig{file=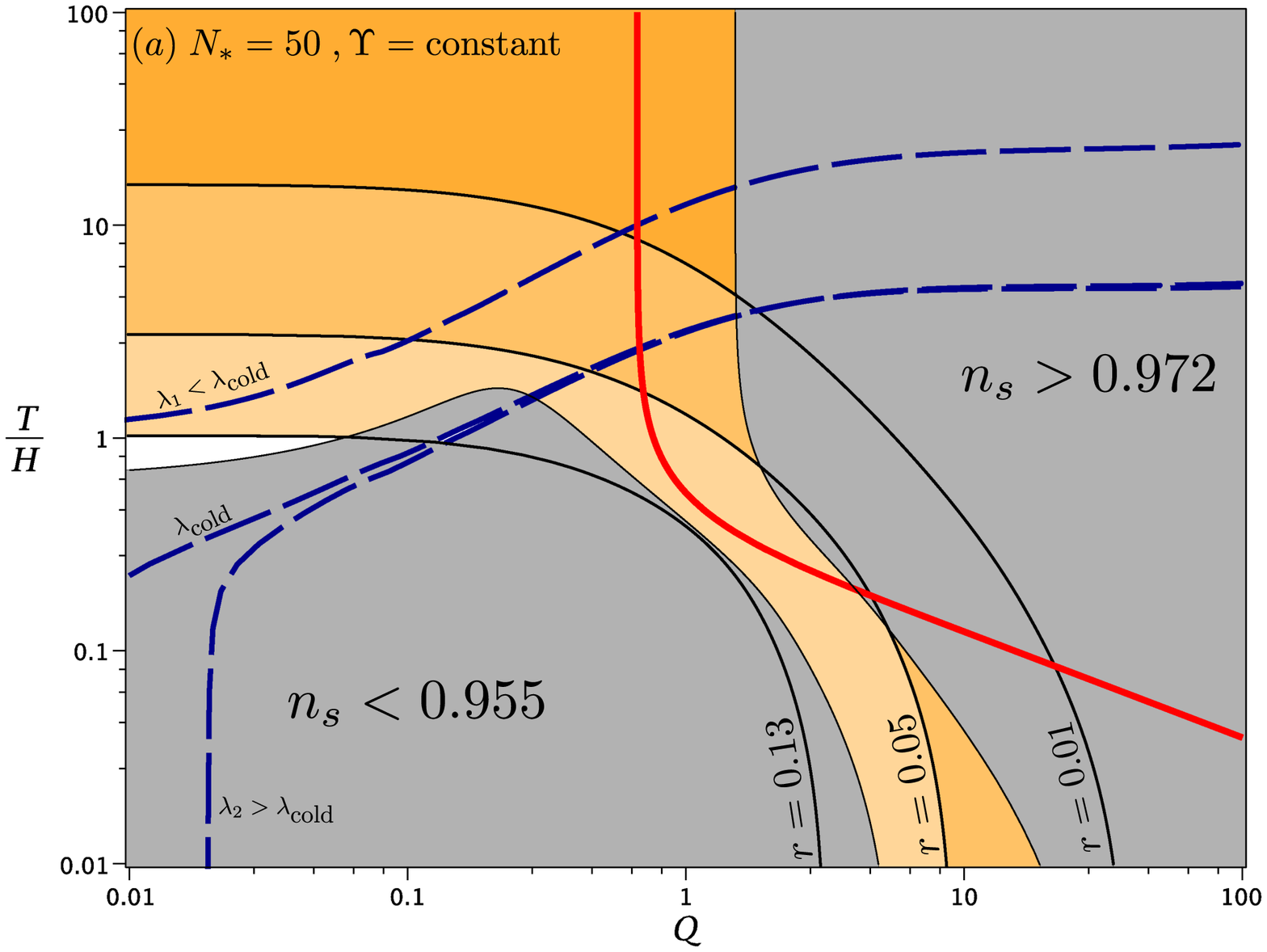,scale=0.4,angle=0}
   \psfig{file=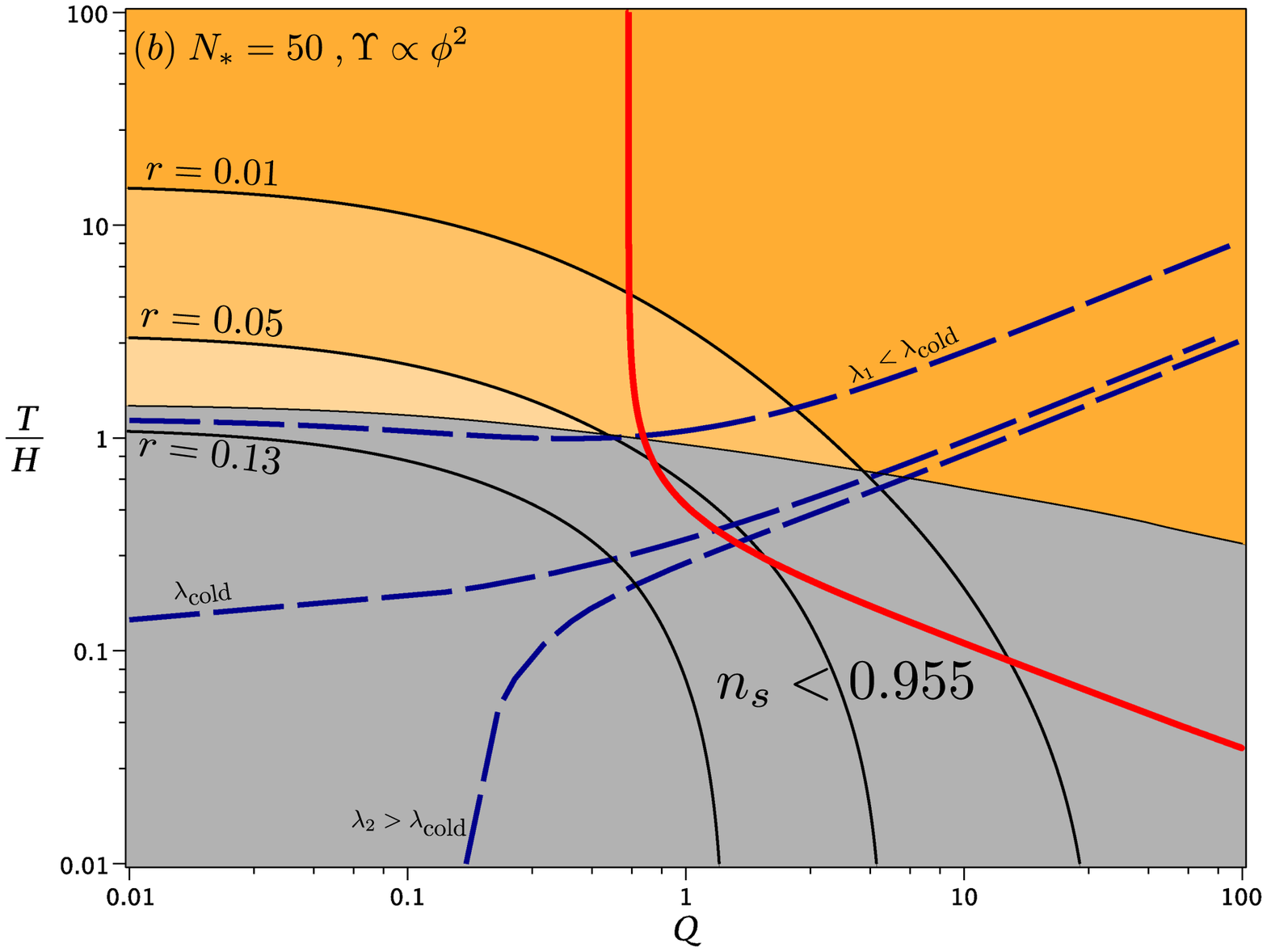,scale=0.4,angle=0}}
\vspace{0.5 cm} \centerline{
  \psfig{file=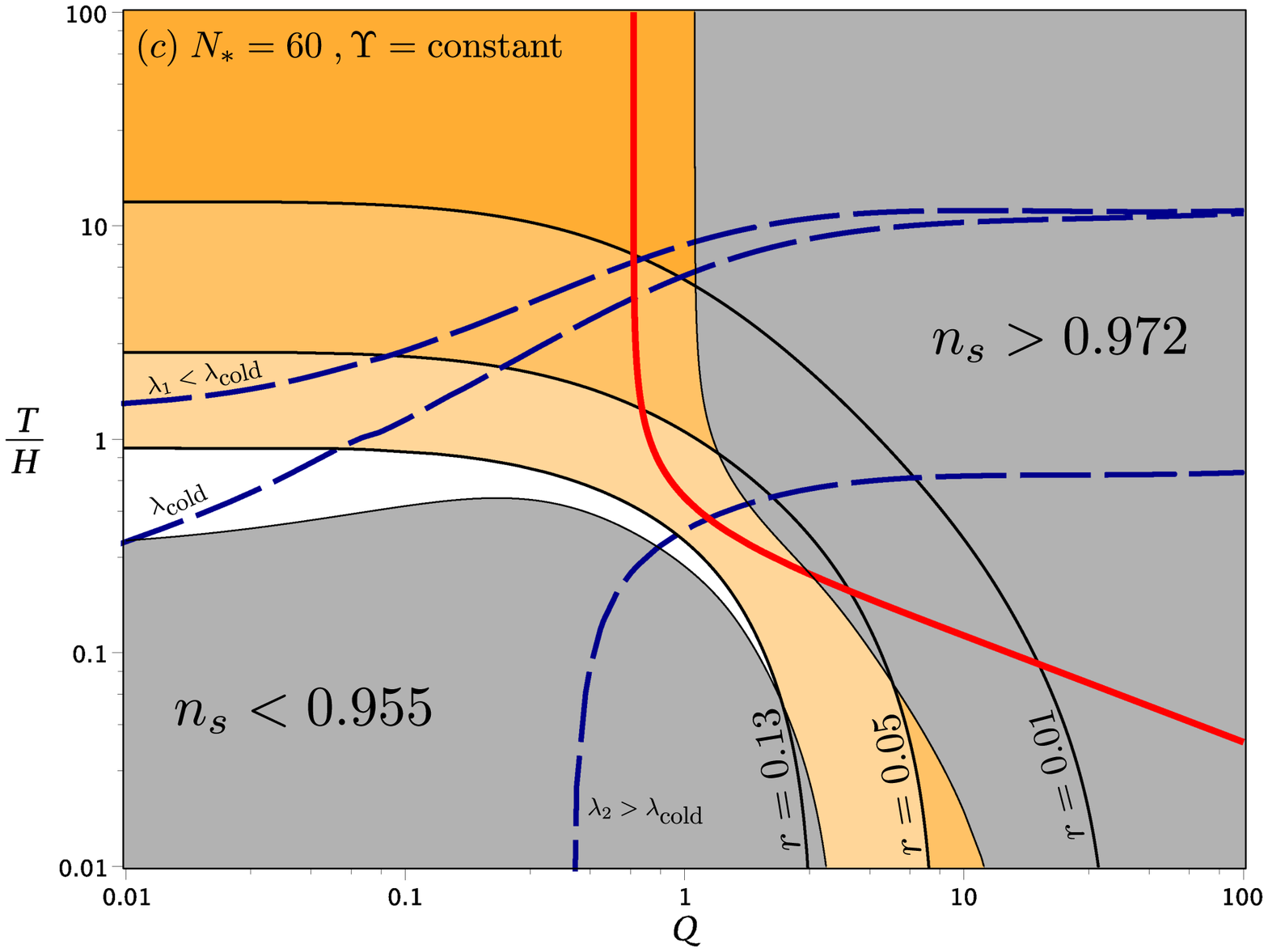,scale=0.4,angle=0}
  \psfig{file=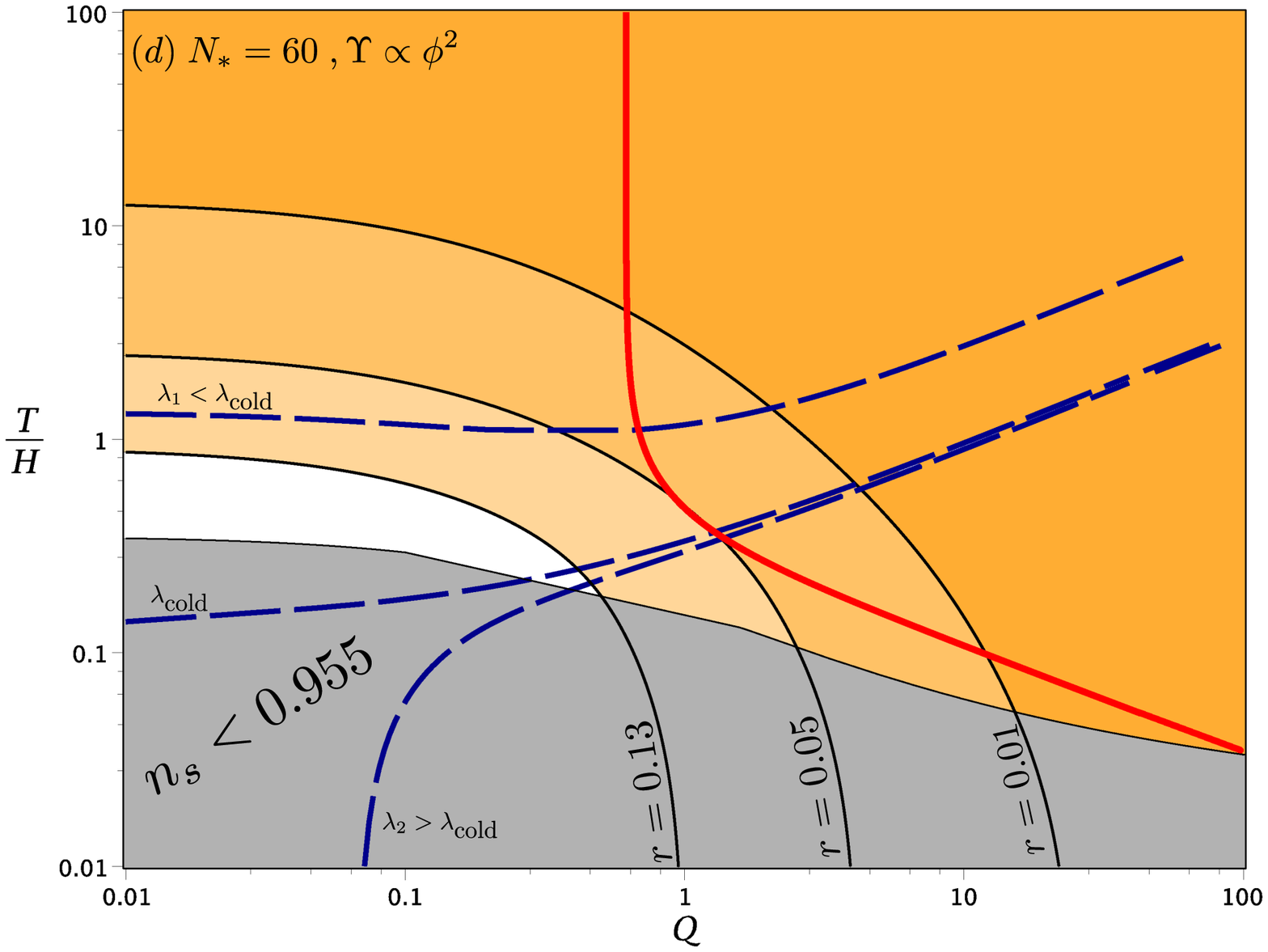,scale=0.4,angle=0}
}
\caption{\sf Results for the $\phi^4$ potential without the thermal
  enhancement factor in the tensor to scalar ratio $r$ expression,
  otherwise, same notation and parameters used in {}figure \ref{phi4}.
  The favorable regions by the observational data are the  orange
  ones.}
   \label{phi4noenhance}
\end{figure}

\begin{figure}[htb]
 \centerline{
   \psfig{file=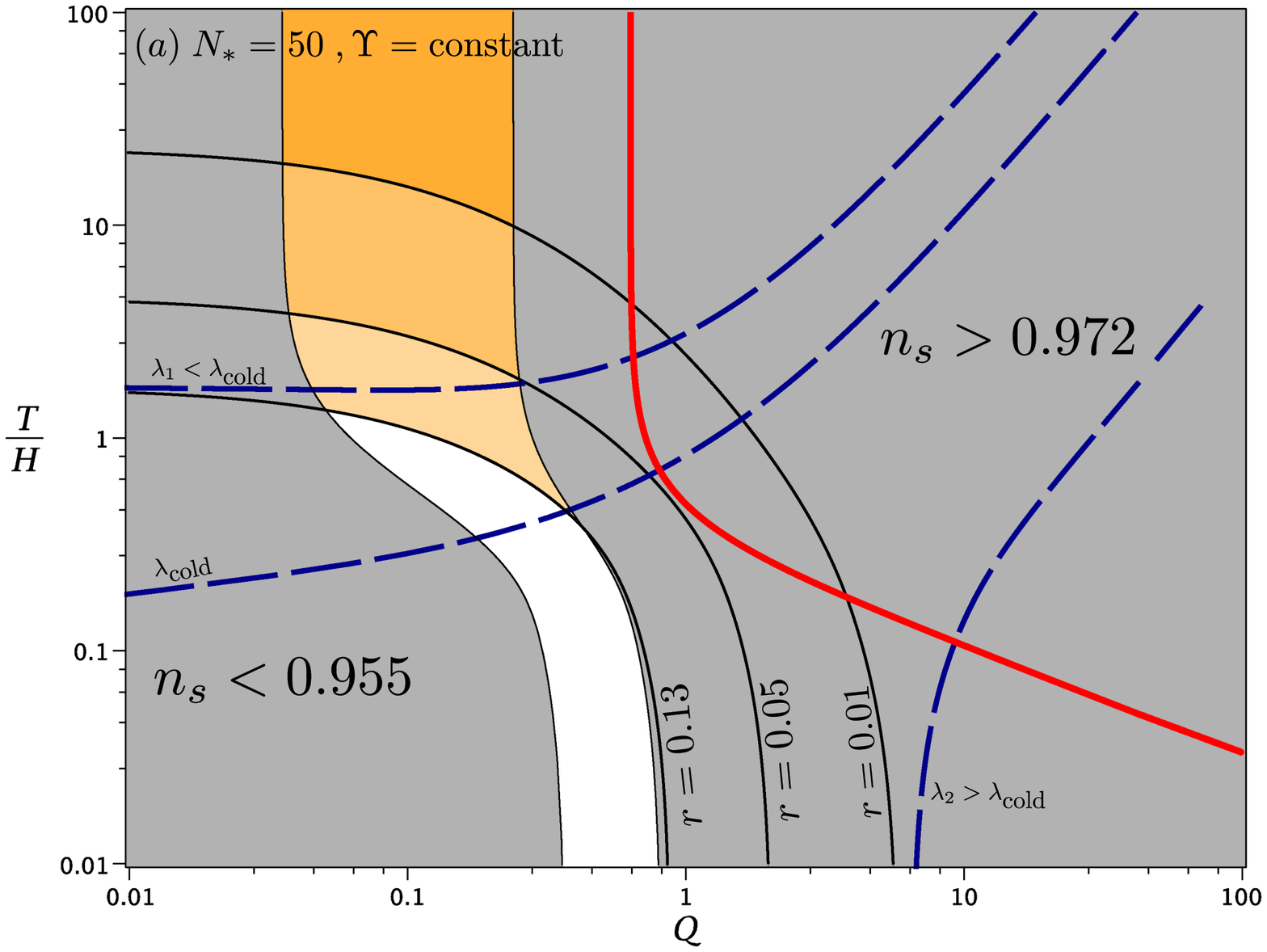,scale=0.4,angle=0}
   \psfig{file=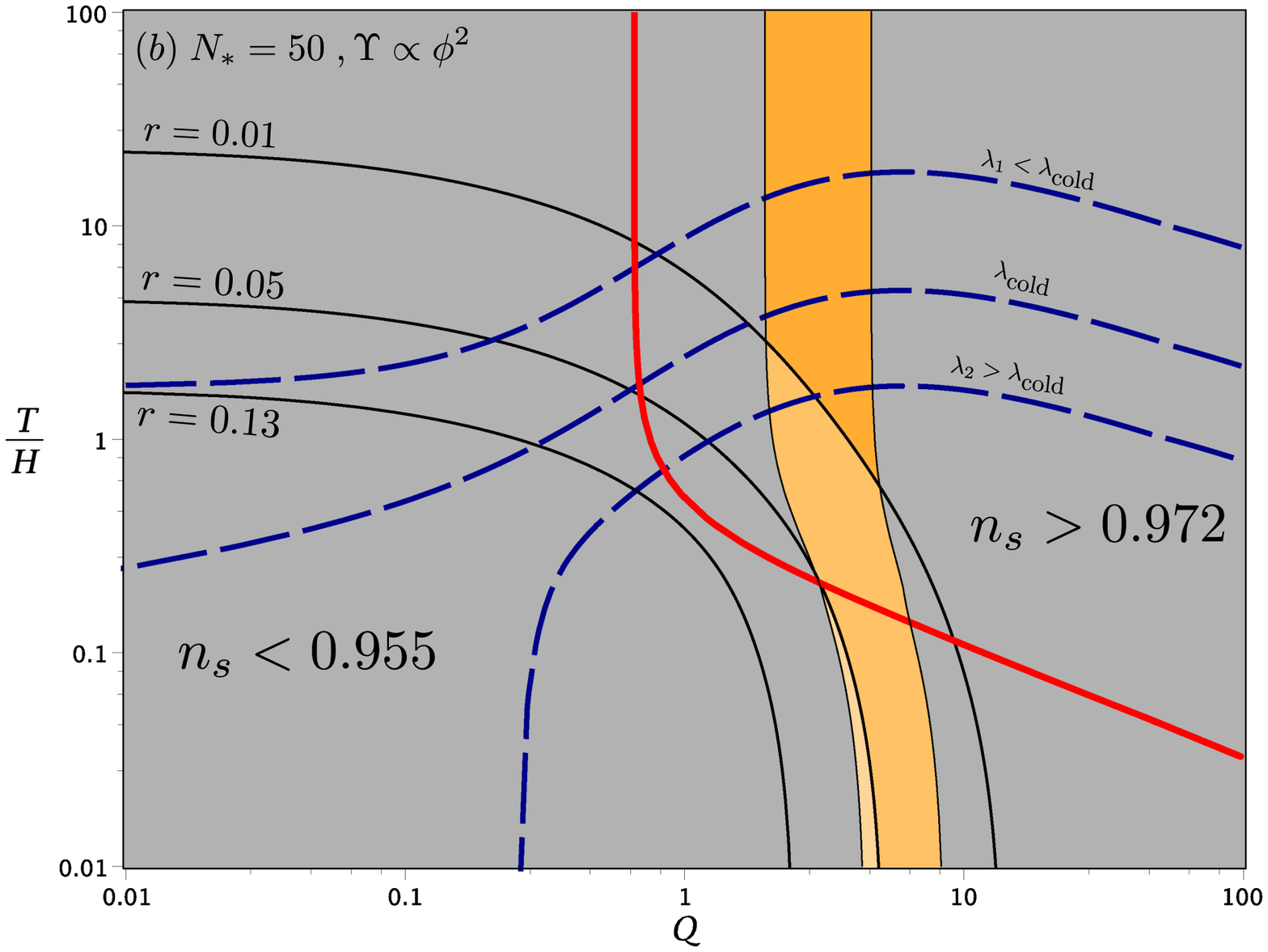,scale=0.4,angle=0}}
\vspace{0.5 cm} \centerline{
  \psfig{file=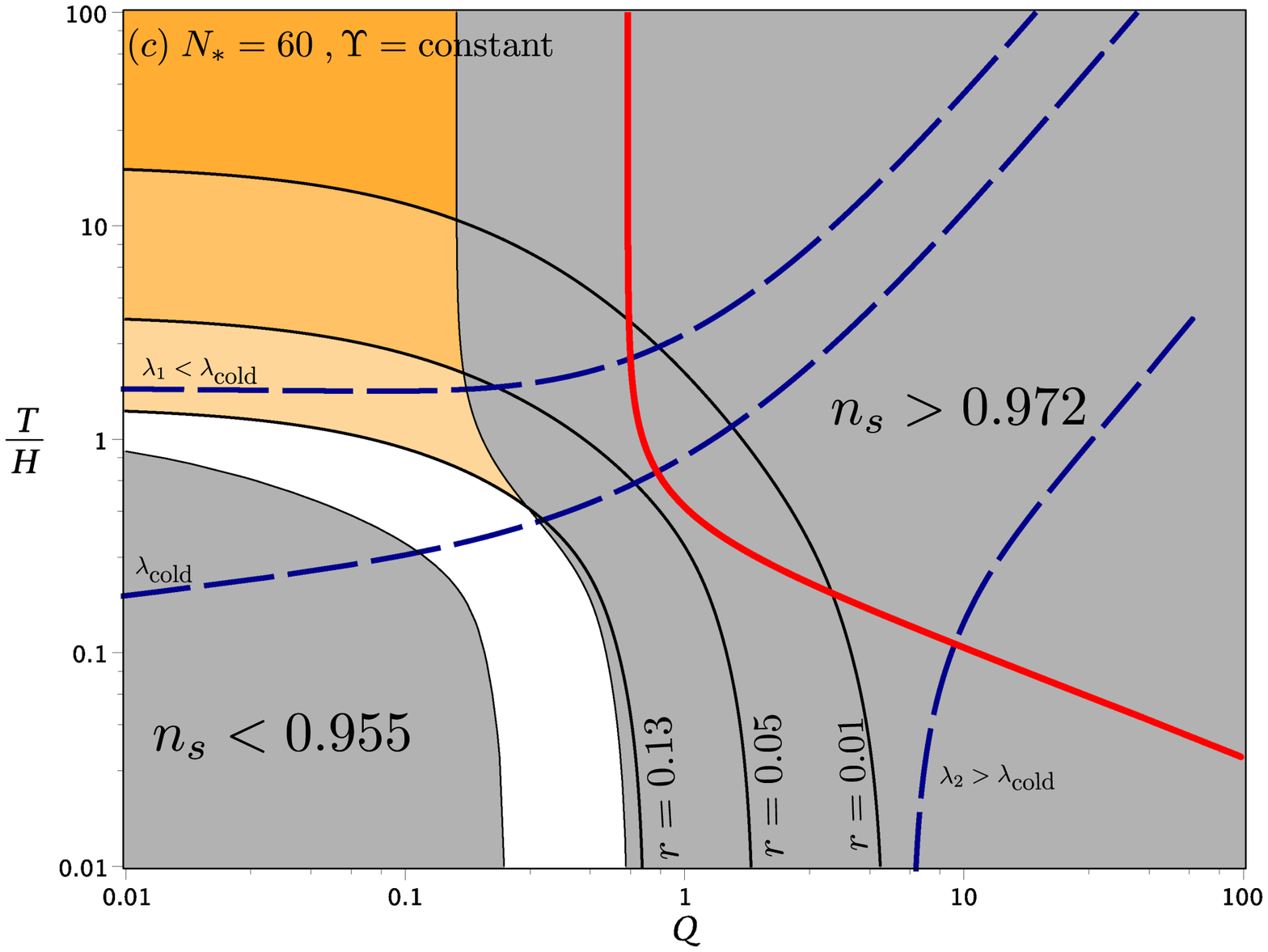,scale=0.4,angle=0}
  \psfig{file=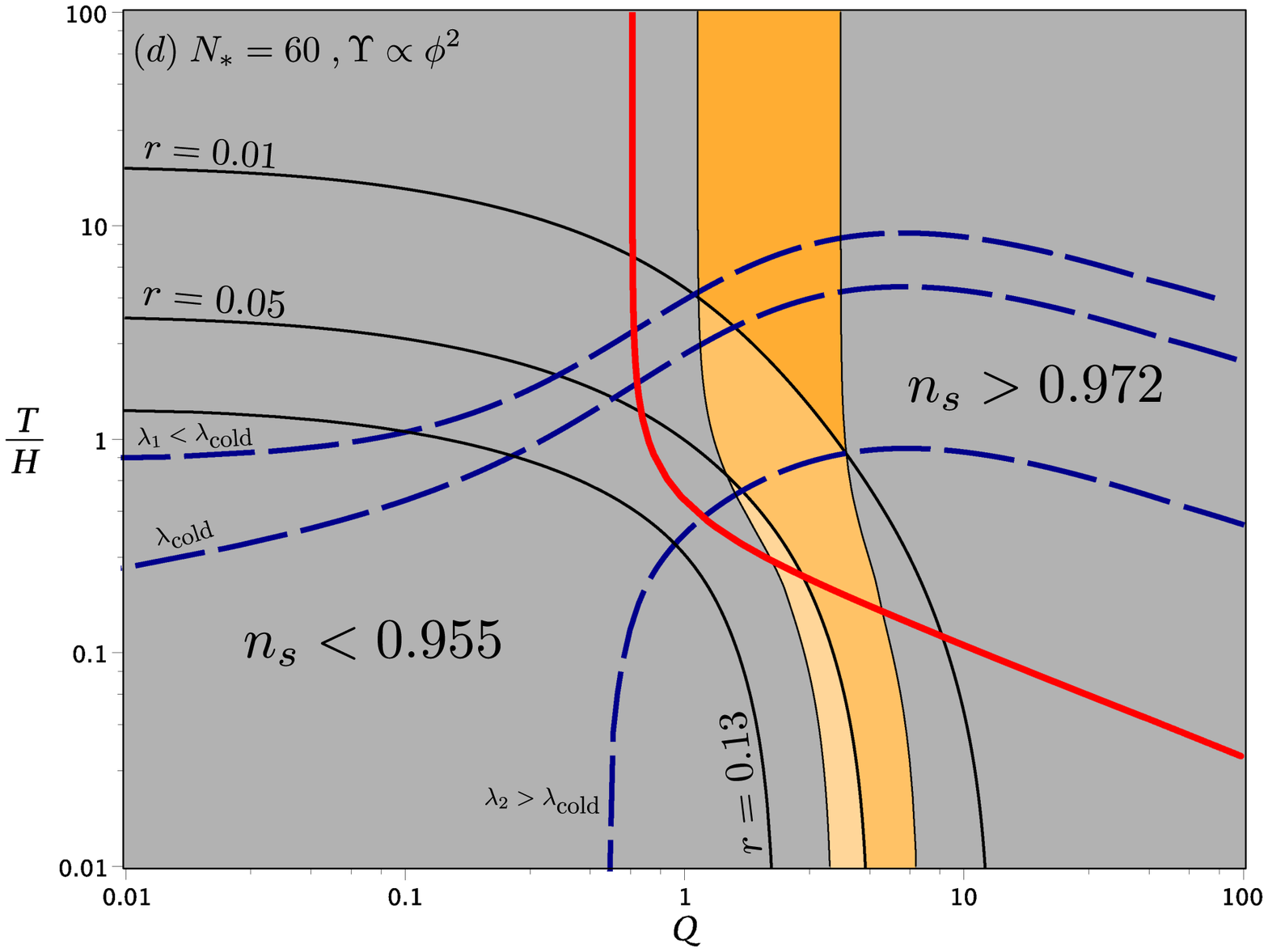,scale=0.4,angle=0}
}
\caption{\sf Results for the $\phi^6$ potential without the thermal
  enhancement factor in the tensor to scalar ratio $r$ expression,
  otherwise, same notation and parameters used in {}figure \ref{phi6}.
  The favorable regions by the observational data are the  orange
  ones.} 
   \label{phi6noenhance}
\end{figure}

Instead of making the more traditional plots in a plane defined by the
confidence levels (95$\%$ CL and 68$\%$ CL) for $n_s$ and $r$ and
specify the inflation models (for different parameters) as points in
this plane, we have chosen, instead, to represent the allowed regions
in the plane $(Q,T/H)$.  In the plane $(Q,T/H)$, we plot for each of
the cases of inflaton potential, $N_*$ and dissipation coefficients,
the permitted regions of parameter space consistent with the
observational values of $n_s$ and $r$. We have also draw  lines of the
inflaton potential parameter $\lambda$ in Eq. (\ref{potential}) that
are consistent with the amplitude of  scalar perturbations
$\Delta_{\cal R}(k_0)$. Different color shades in the plots will
represent  the different regions of parameter space that are allowed
or not.  The obtained results for each case are shown in the {}Figs.
\ref{phi2}, \ref{phi4} and \ref{phi6}, in the case of including the
thermal  enhancement factor in the expression for $r$,
Eq. (\ref{rswi}), while in {}Figs. \ref{phi2noenhance},
\ref{phi4noenhance} and  \ref{phi6noenhance} the thermal enhancement
factor is absent.

In all the Figs. shown in this subsection, we have used the
following representation and notation. The gray and white  areas in
the plots will indicate the regions of parameters  not favored by the
observational data, while the different shades of orange will indicate
those regions of parameters that are favored. The gray regions are the
regions  outside the lower and upper values (from the mean) for the
spectral index, with value taken as $n_s  = 0.9636 \pm 0.0084$,
obtained from the combined data coming from
WMAP9yr+eCMB+BAO+$H_0$~\cite{WMAP9yr}.  This value for $n_s$ is within
the 68$\%$ CL region.  The white areas are for $r> 0.13$
(corresponding  to a region disfavored by the observational data at
the 95$\%$ CL), for the cases  of the quartic and sextic inflaton
potentials, or for $r>0.08$ (corresponding  to a region disfavored by
the observational data at the 68$\%$ CL), for the case of the
quadratic inflaton potential.  The different shades of orange, from
light to dark, separated by thin black lines, indicate the regions of
parameters that have decreasing values for the tensor to scalar
curvature spectrum ratio $r$.  {}For the  the quadratic  inflaton
potential ($p=2$), we have $0.04 < r < 0.08$ (light orange), $0.04 < r
< 0.01$ (medium orange) and $r<0.01$ (dark orange),
respectively. {}For the quartic ($p=4$) and sextic ($p=6$) inflaton
potentials, we have  $0.05 < r < 0.13$ (light orange), $0.05 < r <
0.01$ (medium orange) and $r<0.01$ (dark orange),  respectively.
Results for the amplitude of scalar perturbations are given by the
long-dashed thick blue lines.  These long-dashed thick blue lines are
for three different values of the parameter $\lambda$ in
Eq.~(\ref{potential}) that are consistent with the central value for
the amplitude of  scalar perturbations, $\Delta_{\cal R}^2 = 2.41
\times 10^{-9}$ (WMAP9yr+eCMB+BAO+$H_0$). $\lambda_{\rm cold}$ is the
value required by the pure cold inflation case ($Q=0$ and $T=0$),
$\lambda_1$ and $\lambda_2$ are values below and above $\lambda_{\rm
  cold}$, respectively. All numerical values used  for $\lambda$ are
indicated in the figure captions for each case.  {}Finally, the thick
red line in each of the plots separate the regions dominated by
quantum fluctuations (region below the line), or dominated by thermal
fluctuations (region above the line), according to
{}figure~\ref{PqoverPt}.

In the {}figure \ref{phi2} we have the results for the quadratic
inflaton potential ($p=2$).  In the cold inflation regime, for both
$N_{*}=50$ and $N_*=60$,  this potential is already ruled out by
observations (at 68$\%$ CL)~\cite{WMAP9yr} in the pure cold inflation
case ($Q=0$, $T=0$) and it is a viable potential in this case only at
the 95$\%$ CL.  However, for nonisentropic inflation,  compatibility
with observational data is recovered. {}For example, in the case of
constant dissipation, compatible regions appear for low temperatures,
$T \lesssim H$, and  strong dissipation, $Q \gtrsim 1$. In the case of
inflaton field dependent dissipation,  favored regions can run for all
values of temperature, but for restrict values of dissipation, $0.1
\lesssim Q \lesssim 7.5$ (for $N_* =60$), or  $0.1 \lesssim Q \lesssim
2.5$ (for $N_* =50$).  Regions of thermal fluctuation domination are
more favored in the inflaton field dependent dissipation case only,
while for the constant dissipation case quantum fluctuation tend to
give the most dominant contribution only.

In the {}figure \ref{phi4} we have the results for the quartic
inflaton potential ($p=4$).  This potential is ruled out in the pure
cold inflation case ($Q=0$, $T=0$), for both the $N_{*}=50$ and
$N_{*}=60$ cases, already at 95$\%$ CL when using the recent combined
CMB data from  from WMAP9yr+eCMB+BAO+$H_0$~\cite{WMAP9yr}.  However,
just like in the $p=2$ case for nonisentropic inflation,  this
potential can also become  again compatible with the observational
data. All panels in {}figure~\ref{phi4} show compatible regions with
quantum and thermal fluctuations domination.  However, thermal
fluctuations (warm inflation) dominated regions are more favorable  in
the field dependent dissipation cases, panels (b) and (d). The cases
of constant dissipation, panels (a) and (c), show, however, much
smaller compatible regions with the observational data and  that it
happens only in a thin strip region of $Q$ values ranging from
somewhere in the range $1 \lesssim Q \lesssim 20$, at least in the
presence of the thermal enhancement factor in Eq.~(\ref{rswi}) (see
the results shown below, in the case  when  the enhancement factor in
$r$ is not included).

In the {}figure \ref{phi6} we have the results for the sextic inflaton
potential ($p=6$).  This potential is completely ruled out in the pure
cold inflation case ($Q=0$, $T=0$), for both the $N_{*}=50$ and
$N_{*}=60$ cases, even when taking WMAP only
results~\cite{WMAP9yr}. This inflaton potential is the most
constrained one by the observational data even in the nonisentropic
inflation case.  In particular, in the case of constant dissipation
(panels (a) and (c) in {}figure \ref{phi6}),  it is discarded by not
having regions compatible with the upper bound on $r$, $r <0.13$ at
95$\%$ CL (again, just like in the case of the quartic potential, this
situation can change when discarding the enhancement factor in $r$, as
we will discuss below). In the case of a dissipation  coefficient that
depends on the inflaton field (panels (b) and (d) in {}figure
\ref{phi6}), a compatible region with the observational data only
appears for a small strip of values of $Q$, ranging in between $1
\lesssim Q \lesssim 10$. Inside this range of $Q$, we find compatible
regions for all values of temperature and also covering both quantum
fluctuation dominated regions (that happens for $T/H \lesssim 0.1$)
and thermal fluctuation dominated regions (for $T/H \gtrsim 0.1$).

{}For the sake of comparison, in {}Figs. \ref{phi2noenhance},
\ref{phi4noenhance} and \ref{phi6noenhance} we show the parameter
space behavior when we do not take into account the thermal
enhancement term in Eq.~(\ref{rswi}), i.e., when considering a
spectrum of gravitational waves that is decoupled from the thermal
radiation bath generated during nonisentropic inflation.  Note now
that for all cases the region of parameters consistent with the
observations increase, allowing as well larger regions for which  $r<
0.01$ (dark orange shaded regions). These results show that the cases
of constant dissipation tend to favor more the quantum fluctuation
dominated regime, with very small or completely absent regions of
thermal fluctuation dominated regime (like for a sextic inflaton
potential, $p=6$), while for a dissipation coefficient that depends on
the inflaton field, we can have regions of parameters in both
regimes.  Note also that the larger region of favorable parameters
appear when  considering the quartic inflaton potential, with the
sextic potential  still being the most constrained case.

\subsection{Planck results}

As a final note, we notice that the recently released Planck 
results~\cite{Ade:2013rta}
give for the tensor to scalar amplitude ratio the result
$r < 0.11$ at $95\%$ CL (when the high-$\ell$ CMB
ACT+SPT data are added) and for the spectral index $n_s = 0.9600 \pm 0.0072$,
while when including the Planck lensing likelihood gives 
$n_s = 0.9653 \pm 0.0069$ and $r < 0.13$, and by also adding
BAO data, it gives $n_s = 0.9643 \pm 0.0059$ and $r < 0.12$.
These values are to be compared with the ones we have used here,
$n_s  = 0.9636 \pm 0.0084$ and $r<0.13$ (at $95\%$ CL),
obtained from the combined data coming from
WMAP9yr+eCMB+BAO+$H_0$~\cite{WMAP9yr}, which is quite consistent with those
results from Planck.
Therefore, all our results obtained previously remain fully valid
when also accounting for the recent Planck data,
indicating that the pressure on single field chaotic polynomial inflation models
observed for cold inflation, does not hold in general for 
nonisentropic inflation models.

\section{Conclusions}
\label{sec6}

With the increasing precision that are reaching the cosmological
parameters obtained from the observational data,  considerable
constraints are been put on many inflationary models.  With the recent
release of the accumulated nine years of WMAP data, combined with
other CMB data from ACT and SPT and other astrophysical data coming
from BAO and precise measurements of $H_0$ from large scale surveys,
the simplest single field polynomial potential chaotic  inflation
models with quadratic and higher powers are ruled out at the 95$\%$
CL. Even simple monotonic inflation models appear disfavored at the
68$\%$ CL.  These results, however, only apply to pure cold inflation
models, in their most simple implementations and where the
perturbation spectra is generated by quantum fluctuations only.  In
this work, we have analyzed a much larger class of inflation models,
the nonisentropic inflation models. In these type of models the
perturbation spectra is not sole determined by quantum fluctuations of
the inflaton field, but because in this class of models a non
negligible amount of radiation can be produced during inflation,
thermal fluctuations can as well make a significant contribution to
the power spectra. An example of a nonisentropic inflation model is
warm inflation, where perturbation spectra come dominantly from
thermal fluctuations. Warm inflation models are known to have a much
less constrained  parameter space and these models can be made
compatible with the observational data. However, no study has been
made to date in order to explore parameter regimes ranging from pure
cold inflation to warm inflation. We have made this analysis in this
work.

By appropriately making use of the stochastic inflation program, we
have successfully been able to describe both cold and warm inflation
concomitantly. This has allowed us to obtain the combined
contributions to the power spectrum coming from both quantum and
thermal fluctuations. {}From these results, we were able to fully
determine regions of parameter space when quantum fluctuations give
the dominant contribution, or the opposite regime, when thermal
fluctuations dominate the spectrum.

Since the observational data is strongly constraining and also putting
considerable pressure on polynomial chaotic inflation type of models,
which are also the type of models more easily prone to a microscopic
quantum field theory description, as far model building for inflation
is concerned, we have concentrated our analysis on these type of
inflation models.  We have, thus, investigated the viability of the
nonisentropic polynomial chaotic inflation models in face of the most
recent results for the cosmological parameters, including the
amplitude of scalar curvature perturbations, the spectral index for
the scalar perturbations and the ratio of tensor to scalar
perturbations.  We have investigated the cases of a quadratic, quartic
and a sextic inflaton potential.  In all cases we have found that the
presence of a radiation bath and dissipation in the inflaton dynamics
(which are the characteristics of all nonisentropic inflation models)
can lead to compatible regions of parameter space with the
observational data, re-establishing the viability of these inflaton
potential models.          

Many characteristics of nonisentropic inflation models can be strongly
dependent on the quantum field microphysics dynamics involved during
inflation.  These include, for example, the details of the model
building, like how exactly the  radiation bath is produced, the
thermalization of the inflaton perturbations and the  magnitude and
dependence of the dissipation terms present during inflation. Even so,
the approach we have adopted here is mostly model independent and our
results can easily be extended to other type of inflation models,
including, for example,  hybrid inflation models, models motivated by
string and superstring  theories~\cite{Cai:2010wt,Cerezo:2012ub}, loop
quantum gravity motivated inflation  models~\cite{Herrera:2010yg},
braneworld type of
models~\cite{Nozari:2009th,BasteroGil:2011mr,Herrera:2011zz},
inflation models with nonstandard potentials~\cite{Herrera:2012ep},
among many others.


\appendix
\section{Evaluation of the spectral index $n_s$}

In order to determine the expression for $n_s$,  it is useful to
define the following derivatives in terms of the slow-roll parameters:

\begin{align} \label{ddotphit}
\left. \dfrac{1}{H} \dfrac{d}{dt} \ln \dot \phi \right|_{k=aH} &=
\frac{1}{\Lambda}\left[-3c(1+Q)\delta -
  \frac{c(1+Q)-4}{1+Q}\varepsilon + (c-4)\eta + \frac{4Q}{1+Q}\beta
  \right]\;,
\end{align}
where $\Lambda = 4(1+Q) + (Q-1)c$. {}From the Friedmann equation for
$H$, we obtain

\begin{equation} \label{dHt}
\left. \dfrac{1}{H} \dfrac{d}{dt} \ln H \right|_{k=aH} = -
\frac{1}{1+Q}\varepsilon \;.
\end{equation}
Another useful expression is 

\begin{equation} \label{dst}
\left.\frac{1}{H} \frac{d \ln s}{dt}\right|_{k=aH} = \frac{1}{\Lambda}
\left[ \frac{3(cQ + Q + 1 -c)(1+Q)}{Q}\delta +
  \frac{3Q+9}{1+Q}\varepsilon - 6\eta + \frac{3(Q-1)}{1+Q}\beta
  \right]  \;.
\end{equation}

{}For a given mode with wavenumber $k$, we can then write

\begin{equation} \label{dkt}
\left.\frac{d k}{dt}\right|_{k=aH} = aH^2\left[1 + \dfrac{1}{H}
  \dfrac{d}{dt} \ln H   \right] \;.
\end{equation}
Using also the equation for the entropy density in the slow-roll
regime,

\begin{equation}
Ts = Q \dot \phi^2\;,
\end{equation}
then, $\ln{T} + \ln{s} = \ln{Q} + 2\ln{\dot \phi}$.  Using also
Eq. ({\ref{dst}}) combined with the previous ones, we obtain the
relation

\begin{eqnarray} \label{dTt}
\frac{1}{H} \frac{d}{dt} \ln{T} &=& -\frac{1}{H} \frac{d}{dt} \ln{s} +
\frac{2}{H} \frac{d}{dt} \ln{\dot \phi} +  \frac{1}{H} \frac{d}{dt}
\ln{Q} \nonumber \\ & =& -\frac{1}{H} \frac{d}{dt} \ln{s} +
\frac{2}{H} \frac{d}{dt} \ln{\dot \phi} -  \frac{1}{H} \frac{d}{dt}
\ln{H} +\frac{1}{H\Upsilon}\frac{d\Upsilon}{dk}\frac{dk}{dt}\;.
\end{eqnarray}

Now we have to work out the $d\Upsilon/dk$ term. Assuming $\Upsilon
\equiv \Upsilon(\phi,T)$, then

\begin{eqnarray}
\frac{d\Upsilon}{dk} &=&  \frac{\partial \Upsilon}{\partial \phi}
\frac{d\phi}{dk} +  \frac{\partial \Upsilon}{\partial T} \frac{dT}{dk}
\nonumber \\ &=& \Upsilon_{,\phi} \dot \phi \frac{d t}{dk}  +
\Upsilon_{,T} HT \frac{d t}{dk} \left( \frac{1}{H} \frac{d}{dt} \ln{T}
\right)\;.
\end{eqnarray}

Using Eq. (\ref{dTt}), we obtain:

\begin{eqnarray}
\frac{d\Upsilon}{dk} &=& \Upsilon_{,\phi} \dot \phi \frac{d t}{dk}  +
\Upsilon_{,T} HT  \left[ -\frac{1}{H} \frac{d}{dt} \ln{s} +
  \frac{2}{H} \frac{d}{dt} \ln{\dot \phi} -  \frac{1}{H} \frac{d}{dt}
  \ln{H} \right] \frac{dt}{dk} + \frac{\Upsilon_{,T}T}{\Upsilon}
\frac{d\Upsilon}{dk}\;.
\end{eqnarray}

Using $c \equiv \Upsilon_{,T}T/\Upsilon$, we obtain:

\begin{equation} 
\frac{d\Upsilon}{dk} =\frac{\Upsilon_{,\phi} \dot \phi}{(1-c)} \frac{d
  t}{dk}  + \frac{c}{(1-c)} \Upsilon H  \left[ -\frac{1}{H}
  \frac{d}{dt} \ln{s} + \frac{2}{H} \frac{d}{dt} \ln{\dot \phi} -
  \frac{1}{H} \frac{d}{dt} \ln{H} \right] \frac{dt}{dk}\;. 
\end{equation}

Using $\dfrac{\Upsilon_{,\phi} \dot \phi}{H^2} = -\dfrac{3Q}{(1+Q)}
\beta $ and $\Upsilon H = 3H^2 Q$, we obtain from the above equation,

\begin{equation} \label{dUpsdk}
\left. \frac{d\Upsilon}{dk}\right|_{k=aH} = 3Q \left(H^2
\left.\frac{dt}{dk}\right|_{k=aH}\right) \left\{
-\frac{\beta}{(1+Q)(1-c)}  + \frac{c}{(1-c)} \left[ -\frac{1}{H}
  \frac{d}{dt} \ln{s} + \frac{2}{H} \frac{d}{dt} \ln{\dot \phi} -
  \frac{1}{H} \frac{d}{dt} \ln{H} \right]\right\}\;. 
\end{equation}

If we want to evaluate Eq.~(\ref{dTt}) in terms of slow-roll
parameters, $c$ and $dk/dt$, Eq.~(\ref{dkt}), we simply insert
Eq.~(\ref{dUpsdk}) in Eq.~(\ref{dTt}). Note that for the cases of no
temperature dependence on the dissipation coefficient, then we would
have $c=0$ in the above expressions. Likewise, for a  constant
dissipation, we would have $\beta=0$. Analogously, we have:

\begin{equation}
\left. \frac{d H}{d k}\right|_{k=aH} = H^2
\left.\frac{dt}{dk}\right|_{k=aH} \left(\frac{1}{H} \frac{d}{dt}
\ln{H}\right) \;,
\end{equation}

and

\begin{equation}
\left. \frac{d T}{d k}\right|_{k=aH} =\left. \left\{ H T \frac{dt}{dk}
\left[  -\frac{1}{H} \frac{d}{dt} \ln{s} + \frac{2}{H} \frac{d}{dt}
  \ln{\dot \phi} -  \frac{1}{H} \frac{d}{dt} \ln{H} \right] +
\frac{T}{3HQ} \frac{d\Upsilon}{dk} \right\}\right|_{k=aH}   \;.
\end{equation}


\acknowledgments

R.O.R. is partially supported by research grants from Conselho
Nacional de Desenvolvimento Cient\'{\i}fico e Tecnol\'ogico (CNPq) and
Funda\c{c}\~ao Carlos Chagas Filho de Amparo \`a Pesquisa do Estado do
Rio de Janeiro (FAPERJ). L.A.S. was supported by Coordena\c{c}\~ao de
Aperfei\c{c}oamento de Pessoal de N\'{\i}vel Superior (CAPES).


\end{document}